\documentclass[a4paper,10pt]{article}
\usepackage{graphicx}
\usepackage{amsmath}
\usepackage{citesort}
\usepackage{amsfonts}
\usepackage{amssymb}
\usepackage{amsmath}
\usepackage{latexsym}
\usepackage{color}
\usepackage{ntheorem}
\usepackage{braket}
\usepackage{pstricks}
\usepackage{psfrag}

\newcommand{\ub}[1]{\underbrace{#1}}

\newcommand{\twogL}{\stackrel{\tiny (2)}{g}}

\newcommand{\Nnmo}{N^{n-1}}

\newcommand{\jlcax}[1]{}
\newcommand{\eean}{\nonumber\end{eqnarray}}

\newcommand{\Uone}{\mathrm{U(1) }}%

\newcommand{\kk}[1]{}

\newcommand{\beq}{\begin{equation}}

\newcommand{\FS}       
                  {F}

\newcommand{\HS} 
       {H_{\mbox{\scriptsize volume}}}

{\ptc{this should be removed in the oberwolfach version}}%

\newcommand{\mcA}{{\mycal A}}%
\newcommand{\eeal}[1]{\label{#1}\end{eqnarray}}
\newcommand{\bed}{\begin{deqarr}}
\newcommand{\eed}{\end{deqarr}}
\newcommand{\bedl}[1]{\begin{deqarr}\label{#1}}
\newcommand{\eedl}[2]{\arrlabel{#1}\label{#2}\end{deqarr}}

\newcommand{\mcU}{{\mycal U}}

\newcommand{\bel}[1]{\begin{equation}\label{#1}}
\newcommand{\bea}{\begin{eqnarray}}
\newcommand{\bean}{\begin{eqnarray}\nonumber}
\newcommand{\beal}[1]{\begin{eqnarray}\label{#1}}
\newcommand{\eea}{\end{eqnarray}}

\newcommand{\Eq}[1]{Equation~\eq{#1}}

\newcommand{\qed}{\hfill $\Box$ \medskip}

\newcommand{\be}{\begin{equation}}
\newcommand{\eeq}{\end{equation}}
\newcommand{\ee}{\end{equation}}
\newcommand{\beqa}{\begin{eqnarray}}
\newcommand{\eeqa}{\end{eqnarray}}
\newcommand{\beqan}{\begin{eqnarray*}}
\newcommand{\eeqan}{\end{eqnarray*}}
\newcommand{\ba}{\begin{array}}
\newcommand{\ea}{\end{array}}

\newcommand{\const}{\mbox{\rm const}} 

\newcommand{\mcM}{{\mycal M}}

\newcommand{\mcW}{{\mycal W}}
\newcommand{\mcV}{{\mycal V}}

\newcommand{\scri}{{\mycal I}}%
\newcommand{\scrip}{\scri^{+}}%
\newcommand{\scrim}{\scri^{-}}%
\newcommand{\scripm}{\scri^{\pm}}%

\newcommand{\mnote}[1]
{\protect{\stepcounter{mnotecount}}$^{\mbox{\footnotesize
$
\bullet$\themnotecount}}$ \marginpar{
\raggedright\tiny\em
$\!\!\!\!\!\!\,\bullet$\themnotecount: #1} }

\newcommand{\warn}[1]
{\protect{\stepcounter{mnotecount}}$^{\mbox{\footnotesize
$
\bullet$\themnotecount}}$ \marginpar{
\raggedright\tiny\em
$\!\!\!\!\!\!\,\bullet$\themnotecount: {\bf Warning:} #1} }

\newcommand{\R}{\mathbb R}

\newcommand{\Z}{\mathbb Z}

\newcommand{\eq}[1]{(\ref{#1})}

\newcommand{\ptc}[1]{\mnote{{\bf ptc:}#1}}

\newcommand{\beqar}{\begin{deqarr}}
\newcommand{\eeqar}{\end{deqarr}}

\newcommand{\beaa}{\begin{eqnarray*}}
\newcommand{\eeaa}{\end{eqnarray*}}

\newcommand{\hr}{{\hat r}}

\newcommand{\strictlyx}{}

\DeclareFontFamily{OT1}{rsfs}{}
\DeclareFontShape{OT1}{rsfs}{m}{n}{ <-7> rsfs5 <7-10> rsfs7 <10-> rsfs10}{}
\DeclareMathAlphabet{\mycal}{OT1}{rsfs}{m}{n}

{\catcode `\@=11 \global\let\AddToReset=\@addtoreset}
\AddToReset{equation}{section}

{\catcode `\@=11 \global\let\AddToReset=\@addtoreset}
\AddToReset{figure}{section}

{\catcode `\@=11 \global\let\AddToReset=\@addtoreset}
\AddToReset{table}{section}

\newcounter{mnotecount}[section]
\renewcommand{\themnotecount}{\thesection.\arabic{mnotecount}}
\newtheorem{theorem}{{\sc  Theorem}\rm}[section]

\newtheorem{Remark}[theorem]{{\sc Remark}\rm}

\begin{document}

\title{Space-time diagrammatics%
\thanks{Preprint UWThPh-2012-26}}
\author{
Piotr T. Chru\'sciel\thanks{Email  {
Piotr.Chrusciel@univie.ac.at}, URL {
http://homepage.univie.ac.at/piotr.chrusciel/}}\ \
and Christa R. \"Olz
\\
 University of Vienna
 \\
 \\Sebastian J. Szybka\\Astronomical Observatory, Jagellonian University
}

\date{}
\maketitle
\begin{abstract}
We introduce a new class of two-dimensional diagrams, the \emph{projection diagrams}, as a tool to visualize the global structure of space-times. We construct the diagrams for several metrics of interest, including the Kerr-Newman - (anti) de Sitter family, with or without  cosmological constant, and the Emparan-Reall black rings.
\end{abstract}

\tableofcontents

\section{Introduction}
 \label{S2VII12.3}

A very useful tool for visualizing the geometry of two-dimensional Lorentzian manifolds is that of \emph{conformal Carter-Penrose diagrams}. Such diagrams have been successfully used to visualize the geometry of two-dimensional sections   of the Schwarzschild, (cf., e.g.,~\cite{HE}),   Kerr~\cite{CarterKerr,CarterlesHouches} and several other~\cite{GibbonsHawkingCEH} geometries. A systematic study of conformal diagrams for time-independent two-dimensional geometries has been carried out in~\cite{Walker} by Walker; for the convenience of the reader, Walker's analysis is briefly summarized in Section~\ref{S2VII12.1}.

For spherically symmetric geometries, the two-dimensional conformal diagrams provide useful information about the four-dimensional geometry as well, since many essential aspects of the space-time geometry are  described by
 the $t-r$ sector of the metric.

The object of this paper is to show that one can
 usefully represent classes of non-spherically symmetric geometries in terms of two-dimensional diagrams, which we call \emph{projection diagrams}, using an auxiliary two-dimensional metric, constructed out of the space-time metric. The issues such as stable causality, global hyperbolicity, existence of event or Cauchy horizons, the causal nature of boundaries, and existence of conformally smooth infinities become evident by inspection of the diagrams.

We give a general definition of such diagrams, and construct examples for the Kerr-Newman family of metrics, with or without cosmological constant of either sign, and for the Emparan-Reall metrics. We show how the projection diagrams for the Pomeransky-Senkov metrics could be constructed, and present a tentative diagram for those metrics. We end the paper by pointing out  how the projection diagrams can be used to construct inequivalent extensions of a family of maximal, globally hyperbolic, vacuum or electrovacuum, space-times with compact Cauchy surfaces obtained by periodic identifications of the time coordinate in the Kerr-Newman - (a)dS family of metrics, as well as for Taub-NUT space-times.

\section{Conformal diagrams for static two-dimensional spacetimes}
 \label{S2VII12.1}

Following~\cite{Walker}, we  construct conformal diagrams for two-dimensional Lorentzian metrics of the form
\bea \label{Krakgen2VII12}
 &\twogL=-F(r) dt^{2}+F^{-1}(r)dr^{2} \;,
 &
\eea
where $F$ is, for simplicity and definiteness, a real-analytic  function on   an interval,  $t$ ranges over $\R$, and one considers separately maximal intervals in $\R$ on which $F$ is finite and does not change sign; those define the ranges of $r$. Each such interval leads to a connected Lorentzian manifold on which $\twogL$ is defined, and the issue is whether or not such manifolds can be patched together, and how. Note that $t$ is \emph{not} a time coordinate in regions where $F$ is negative.

It should be kept in mind that the study of the conformal structure for more general metrics of the form
\bea \label{Krakgen2VII12.2}
 &\twogL=-F(r)H_1(r) dt^{2}+F^{-1}(r)H_2(r)dr^{2} \;,
 &
\eea
where $H_1$ and $H_2$ are  \strictlyx
positive  in the range of $r$ of interest, can be reduced to the one
for the metric \eq{Krakgen2VII12} by writing
\bea
 \label{Krakgen2VII12.a2}
 &\twogL=\sqrt{H_1 H_2} \left( -\hat F  dt^{2}+\hat F^{-1} dr^{2}\right) \;,
 \ \mbox{where $\hat F = \sqrt{\frac{H_1}{H_2}}F$}
 \;.
 &
\eea

\subsection{Manifest conformal flatness}
 \label{ss4VII12.1}

\begin{figure}
\begin{center}
{\includegraphics[scale=.7]{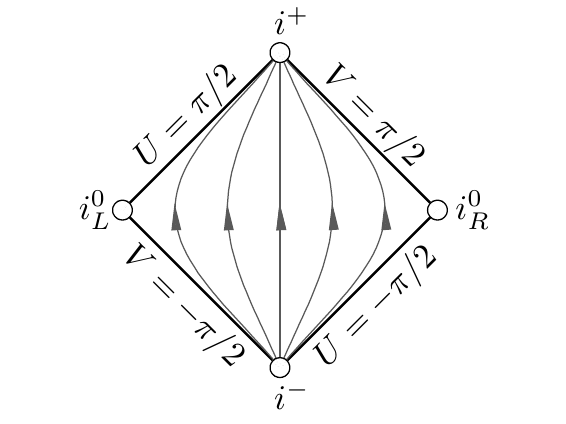}}
\caption{The conformal diagram for $(1+1)$-dimensional Minkowski space-time.%
\label{F2dMc2VII12}}
\end{center}
\end{figure}

In order to bring  the metric \eq{Krakgen2VII12} to a manifestly conformally flat form,  one chooses a value of $r_*$ such that $F(r_*)\ne 0$ and introduces a new coordinate $x$ defined as
\bel{19IV.12VII12}
 x(r)= \int _{r_*}^{r} \frac {ds}{F(s)}
 \quad \Longrightarrow \quad dx = \frac{dr}{F(r)}
 \;,
\ee
leading to
\bel{19IV.22VII12}
 \twogL=-F dt^2 + \frac 1 F ( {F dx})^2 =
  F (-dt^2+dx^2)
 \;.
\ee
The geometry of the space-time, and its possible extendability, will depend upon the sign of $F$, the zeros of $F$, and their order. For example, whenever $x$ ranges over $\R$ the spacetime $(\R^2,\twogL)$ can be conformally mapped to the following \emph{diamond}:
$$
 \{-\pi/2< T-X<\pi/2\;,\quad -\pi/2<T+X<\pi/2
\} \subset \R^2
 \;.
$$
This is done by first introducing
\bel{7IV.4}
 u = t-x\;, \  v= t+x \qquad \Longleftrightarrow
 \qquad t = \frac{u+v} 2\;, \  x = \frac{ v-u} 2
 \;,
\ee
which brings $\twogL$ into the form
$$
\twogL = -du \, dv
\;.
$$
While some other ranges of variables might arise in specific examples, in the current case we have  $(u,v)\in \R^2$.
 We bring the last $\R^2$ to a bounded set using
\bel{7IV.3}
 U = \arctan (u)\;,\qquad V= \arctan (v)
 \;,
\ee
thus
$$
 (U,V) \in \left( - \frac \pi 2, \frac \pi 2\right) \times \left( - \frac \pi 2, \frac \pi 2\right)
 \;.
$$
This looks somewhat more familiar if we make one last change of
coordinates similar to that in \eq{7IV.4}:
\bel{7IV.5}
 U = T-X\;, \  V= T+X \quad \Longleftrightarrow
 \quad T = \frac{U+V} 2\;, \  X = \frac{ V-U} 2
 \;,
\ee
see Figure~\ref{F2dMc2VII12}, leading to
$$
 \twogL = \frac 1 {\cos^2 (T-X) \cos^2 (T+X)} (-dT^2 + dX^2)
 \;.
$$

Simple variations of the above coordinate transformations might be used for alternative ranges of $x$. An integral of $1/F$ which is infinite near one of the integration bounds and finite at the other one leads to triangles, obtained by cutting a diamond across a diagonal; the sign of $F$ determines which diagonal is relevant.
A finite integral of $1/F$ leads to strips, if one does not perform the subsequent coordinate transformation \eq{7IV.3}.
These are then the building blocs, out of which  the final maximal diagrams can be built.

\subsection{Gluing}
 \label{ss4VII12.2}

We pass now to the gluing question. It turns out that
\emph{four blocs} can be  glued together across a boundary $\{r=r_0\}$ at which
$$
 F(r_0)=0\;,\quad F'(r_0)\ne 0\;.
 $$
Since $F$ has a simple zero, it factorizes as $$F(r)=
(r-r_0)H(r)\;,$$ for a function $H$ which has no zeros in a
neighborhood of $r_0$.
The gluing is done in two steps by defining
 \bel{udef22VII12} u=t-f(r)\;,\quad v=t+f(r)\;,\quad f' = \frac 1 {F}\;, \ee
\bel{desingular22VII12} \hat u = -\exp(-c u)\;,\quad \hat v = \exp
(cv)\;,
 \ee
where
$$
 c=\frac{F'(r_0)}{2}
  \;.
$$
This leads to the following form of the metric
\bel{genhatco312VII12} \twogL =\mp \frac{4H(r)}{(F'(r_0))^2} \exp(-\hat
f(r)F'(r_0))d\hat u \, d \hat v  \;, \ee with a negative sign
if we started in the region $r>r_0$, and positive otherwise.
Here
$$   \hat f (r):= f(r) - \frac 1{F'(r_0)} \ln |r-r_0|\;.
$$
In \eq{genhatco312VII12} the function $r$ should be viewed as a function of the product $\hat u \hat v$, implicitly defined by the equation
$$\hat u\hat v = \mp (r-r_0) \exp(\hat f(r)F'(r_0))\;.$$
  Note that for analytic $F$'s the extension so constructed
 is real analytic; this follows from the analytic version of the implicit function
 theorem.

Boundaries at finite distance $r=r_0$ but at which $F$ has a zero of higher order can still be glued together via  \emph{two-blocs gluing}. Here one continues to use  the functions $u$ and $v$ defined
in \eq{udef22VII12}, but now one does \emph{not} use $u$ and $v$ simultaneously as coordinates. Instead one considers a coordinate system $(u,r)$, so
that
$$
 \twogL = -F ( {du+\frac 1 F dr})^2 + \frac 1 F dr^2 = - F du^2 - 2 du\, dr
 \;.
$$
Since $\det \twogL = -1$, the resulting metric extends smoothly
as a Lorentzian metric to the whole nearby interval where $F$ is defined. This will certainly include the nearest conformal bloc, as well as some further ones if the case arises. A distinct extension is obtained when using  instead the coordinate system $(v,r)$.

Asymptotic regions where $|r|\to\infty$ but $F$ is bounded, and bounded away from zero, provide null conformal boundaries at infinity.

\section{Projection diagrams}
 \label{S10VII12.1}

\subsection{The definition}
 \label{S14VIII12.2}

Let $(\mcM,g)$ be a smooth space-time, and let $\R^{1,n}$ denote the $(n+1)$-dimensional Minkowski space-time. A \emph{projection diagram} is a pair $(\pi,\mcU)$, where
\newcommand{\Upsiw}{\mcW}
$$\pi:\mcM\to\Upsiw
$$
is a continuous map, differentiable on an open dense set,
 from $\mcM$ onto $\pi(\mcM)=:\Upsiw\subset \R^{1,1}$; together with
 an open set
$$\mcU\subset \mcM
\;,
$$
assumed to be non-empty, on which $\pi$ is a smooth submersion, so that it holds:
\begin{enumerate}
 \item every smooth timelike curve $\sigma\subset \pi(\mcU) $ is the projection of a smooth timelike curve $\gamma$ in $(\mcU,g)$: $\sigma=\pi\circ\gamma$;
 \item the image $\pi\circ\gamma$ of every smooth timelike curve $\gamma\subset \mcU$ is a timelike curve in $\R^{1,1}$.

\end{enumerate}

Some comments are in order:

First, we have assumed for simplicity that  $(\mcM,g)$, $\pi|_\mcU$, and the causal curves in the definition are smooth, though this is unnecessary for most purposes.

Next, we do not assume that $\pi$ is a submersion, or in fact differentiable, everywhere on $\mcM$. This allows us to talk about ``the projection diagram of Minkowski space-time", or ``the projection diagram of Kerr space-time", rather than of ``the projection diagram of the subset $\mcU$ of Minkowski space-time", etc. Note that the latter terminology would be more precise, and will sometimes be used, but appears to be an overkill in most cases.

Further, the requirement that timelike curves in $\pi(\mcU)$ arise as projections of timelike curves in $\mcM$ ensures that causal relations on  $\pi(\mcU)$, which can be seen by inspection of  $\pi(\mcU)$, reflect causal relations on $\mcM$. Conditions 1 and 2 taken together ensure that causality on  $\pi(\mcU)$ represents as accurately as possible causality on $\mcU$.

By continuity, images of causal curves in $\mcU$ are causal in $\pi(\mcU)$. Note  that null curves in $\mcU$ are often mapped to timelike ones in $\pi(\mcU)$.

The second condition of the definition is of course equivalent to the requirement that the images by $\pi_*$ of timelike vectors in $T\mcU$ are timelike. This implies further that the images by $\pi_*$ of causal vectors in $T\mcU$ are causal. One should keep in mind that images by $\pi_*$ of null vectors in $T\mcU$ could be timelike. And, of course, many spacelike vectors will be mapped to causal vectors under $\pi_*$.

Recall that $\pi$ is a submersion if  $\pi_*$ is surjective at every point. The requirement that $\pi$ is a submersion guarantees that open sets are mapped to open sets. This, in turn, ensures that projection diagrams with the same set $\mcU$ are locally unique, up to a local conformal isometry of two-dimensional Minkowski space-time. We do not know whether or not two  surjective projection diagrams $\pi_i:\mcU\to \Upsiw_i$, $i=1,2$, with identical domain of definition $\mcU$ are globally unique, up to a  conformal isometry of $\Upsiw_1$ and $\Upsiw_2$. It would be of interest to settle this question.

In many examples of interest the set $\mcU$ will \emph{not} be connected.

Note that a necessary condition for existence of a projection diagram is \emph{stable causality} of $\mcU$: indeed, let $t$ be any time function on $\R^{1,1}$, then $t\circ\pi$ is a time function on $\mcU$.

It might be tempting to require that $\mcU$ be dense in $\mcM$. Such a requirement would, however, prohibit one to construct a projection diagram of the usual maximal extension of Kerr space-time, since the latter contains   open regions which are not stably causal.

Recall that a map is proper if inverse images of compact sets are compact. One could further require $\pi$ to be proper; indeed, many projection diagrams below have this property. This is actually useful, as then the inverse images of globally hyperbolic subsets of $\Upsiw$ are globally hyperbolic, and so global hyperbolicity, or lack thereof, can be established by visual inspection of $\Upsiw$. It appears, however, more convenient to talk about \emph{proper projection diagrams} whenever $\pi$ is proper, allowing for non-properness in general.

As such, we have assumed for simplicity that $\pi$ maps $\mcM$
into a subset of Minkowski space-time. In some applications it might be natural to consider more general two-dimensional manifolds as the target of $\pi$; this requires only a trivial modification of the definition. An example is provided by the Gowdy metrics on a torus, discussed at the end of this section,  where the natural image manifold for $\pi$ is $(-\infty,0)\times S^1$, equipped with a flat product metric. Similarly, maximal extensions of the class of Kerr-Newman - de Sitter metrics of Figure~\ref{F25VI12.6}, p.~\pageref{F25VI12.6}, require the image of $\pi$ to be a suitable Riemann surface.

\subsection{Simplest examples}
 \label{S14VIII12.3}

The simplest examples of projection diagrams can be constructed for metrics of the form
\bea
\label{11VII12.1}
 &g=e^f(-F dt^{2}+F^{-1}dr^{2})+\ub{h_{AB}dx^A
 dx^B}_{=:h}\;,\qquad F=F(r)\;,&
\eea
where $h=h_{AB}(t,r,x^C)dx^A dx^B$ is a family of
\emph{Riemannian} metrics on an $(n-1)$-dimensional manifold
$\Nnmo$, possibly depending upon $t$ and $r$, and $f$ is a function which is allowed to depend upon all variables. It should be clear that any manifestly conformally flat representation
of any extension, defined on  $\Upsiw\subset \R^{1,1}$, of the two-dimensional metric $-Fdt^2 + F^{-1} dr^2$,
 as discussed in Section~\ref{S2VII12.1}, provides immediately a projection diagram for $(\Upsiw\times \Nnmo,g)$.

In particular, introducing spherical coordinates $(t,r,x^A)$ on
\bel{17VIII12.11}
 \mcU:=\{(t,\vec x) \in \R^{n+1}\,, |\vec x|\ne 0\}\subset \R^{1,n}
\ee
and forgetting about the $(n-1)$-sphere-part of the  metric leads to a projection diagram for Minkowski space-time  which coincides with the usual conformal diagram of the fixed-angles subsets of Minkowski space-time (see the left figure in Figure~\ref{F25VII12.98}, p.~\pageref{F25VII12.98} below; the shaded region there should be left unshaded in the Minkowski case).
The set $\mcU$ defined in \eq{17VIII12.11} cannot be extended to include the world-line passing through the origin of $\R^n$ since the map $\pi$ fails to be differentiable there. This diagram is proper, but fails to represent correctly the nature of the space-time near the set $|\vec x|=0$.

On the other hand, a globally defined projection diagram for Minkowski space-time (thus, $(\mcU,g)=\R^{1,n}$) can be obtained by writing $\R^{1,n}$ as a product $\R^{1,1}\times \R^{n-1}$, and forgetting about the second factor. This leads to a projection diagram of Figure~\ref{F2dMc2VII12}. This diagram, which is not proper, fails to represent correctly the connectedness of $\scrip$ and $\scrim$  when $n>1$.

It will be seen in Section~\ref{S2VII12.9} that yet another choice of $\pi$ and of the set $(\mcU,g)\subset \R^{1,n} $ leads to a third projection diagram for Minkowski space-time.

A further  example of non-uniqueness is provided by the projection diagrams for Taub-NUT metrics, discussed in Section~\ref{ss14VIII12.2}.

These examples show that there is no uniqueness in the projection diagrams, and that various such diagrams might carry different information about the causal structure. It is clear that for space-times with intricate causal structure, some information will be lost when projecting to two dimensions. This raises the interesting question,  whether there exists a notion of optimal projection diagram for specific spacetimes. In any case, the examples we give in what follows appear to depict the essential causal properties of the associated space-time, except perhaps for the black ring diagrams of Section~\ref{S2VII12.9}-\ref{S2VII12.10}.

Non-trivial examples of  metrics of the form \eq{11VII12.1} are provided by the Gowdy metrics on a torus~\cite{gowdy71}, which can be written in the form~\cite{gowdy71,ChANoP}
\bel{13VIII12.2}
 g
  =
 e^{f } (-dt^2+d\theta^2)
  +
|t|\left(e^{P }\left( dx^1 + Q \ dx^2\right)^2 +
  e^{-P} (dx^2)^2\right) \;,
\ee
with $t\in (-\infty,0)$ and $(\theta,x^1,x^2)\in S^1 \times S^1 \times S^1$. Unwrapping $\theta$ from $S^1$ to $\R$ and projecting away the $dx^1$ and $dx^2$ factors, one obtains a projection diagram the image of which is the half-space $t<0$ in Minkowski space-time. This can be further compactified as in Section~\ref{ss4VII12.1}, keeping in mind that the asymptotic behavior of the metric for large negative values of $t$~\cite{RingstromGowdyAtPastInfinity} is not compatible with the existence of a smooth conformal completion of the full space-time metric across past null infinity. Note that this projection diagram fails to represent properly the  existence of Cauchy horizons for non-generic~\cite{RingstroemSCC} Gowdy metrics.

Similarly, generic Gowdy metrics on $S^1\times S^2$, $S^3$, or $L(p,q)$ can be written in the form~\cite{gowdy71,ChANoP}
\begin{equation}\label{13VIII12.1}
 g=e^f(-d t^2+d\theta^2)+R_0 \sin( t)\sin(\theta)\left(e^{P }\left( dx^1 + Q \ dx^2\right)^2 +
  e^{-P} (dx^2)^2\right)
  \;,
\end{equation}
with $(t,\theta)\in (0,\pi)\times [0,\pi]$,
leading to the Gowdy square as the projection diagram for the space-time.
(This is the diagram of Figure~\ref{F25VII12.99}, p.~\pageref{F25VII12.99}, where the lower boundary   corresponds to $t=0$, the upper boundary corresponds to $t=\pi$, the left boundary corresponds to the axis of rotation $\theta=0$, and the right boundary is the projection of the axis of rotation $\theta=\pi$. The diagonals, denoted as $y=y_h$ in Figure~\ref{F25VII12.99},  correspond in the Gowdy case to the projection of the set where the gradient of the area  $R=R_0 \sin (t) \sin(\theta) $ of the orbits of the isometry group $\Uone\times\Uone$
 vanishes, and do not have any further geometric significance. The lines with the arrows in Figure~\ref{F25VII12.99} are irrelevant for the Gowdy metrics, as the orbits of the isometry group of the space-time metric are spacelike throughout the Gowdy square.)

In the remainder of this work we will construct projection diagrams for families of metrics of interest which are not of the simple form \eq{11VII12.1}.

\subsection{The Kerr metrics}
 \label{S2VII12.2}

Consider the Kerr metric in Boyer-Lindquist coordinates,
\begin{eqnarray}
 \nonumber
  g &=& - \frac{\Delta_r -a^2 \sin ^2(\theta )}{\Sigma }dt^2
  -\frac{2 a  \sin ^2(\theta ) \left(r^2+a^2-\Delta_r \right)}{\Sigma }dt d\varphi
\\
  &&  +\frac{ \sin ^2(\theta ) \left(\left(r^2+a^2\right)^2-a^2 \sin ^2(\theta )\Delta_r \right)}{\Sigma} d\varphi^2 +\frac{\Sigma }{\Delta_r } dr^2  + \Sigma d\theta^2
  \;.
  \label{Kerr2x}
\end{eqnarray}
Here
\beal{SigDeldef}
 & \Sigma=r^{2}+a^{2}\cos^{2}\theta \;,
\qquad  \Delta_r=r^{2}+a^{2}-2mr= (r-r_+)(r-r_-)
 \;,&
\eea
for some real parameters $a$ and $m$,
with
$$
 r_{\pm}=m\pm(m^{2}-a^{2})^{\frac{1}{2}}
\;, \quad \mbox{and we assume that} \
  0 < |a|\le m
 \;.
$$
We note that
\begin{eqnarray}
  \nonumber
  g_{\varphi\varphi}
    &=& \sin ^2(\theta ) \left(\frac{2 a^2 m r \sin ^2(\theta )}{a^2 \cos ^2(\theta )+r^2}+a^2+r^2\right) \\
    \label{27VII12.3}
    &=& \frac{\sin ^2(\theta ) \left(a^4+a^2 \cos (2 \theta ) \Delta_r+a^2 r (2 m+3 r)+2 r^4\right)}{a^2 \cos (2 \theta )+a^2+2 r^2}
     \;,
\end{eqnarray}
the first line making clear the non-negativity of $g_{\varphi\varphi}$ for $r\ge 0$.

In the region where $ \partial_\varphi$ is spacelike we rewrite the $t-\varphi$ part of the metric as
\bean
 \lefteqn{
 g_{tt} dt^2 + 2 g_{t\varphi} dt d\varphi + g_{\varphi\varphi} d\varphi^2
}
 &&
\\
 &&
  =  g_{\varphi\varphi} \bigg(d\varphi + \frac {g_{t\varphi}}{g_{\varphi\varphi}} dt\bigg)^2
  + \bigg( g_{tt} - \frac{g_{t\varphi}^2}{g_{\varphi\varphi}} \bigg)dt^2
  \;,
\eeal{22VI12.4}
with
\beaa
 g_{tt} - \frac{g_{t\varphi}^2}{g_{\varphi\varphi}}
   &= &
   -\frac{2 \Delta_r\Sigma  }{a^4+a^2 \Delta_r \cos (2 \theta ) +a^2 r (2 m+3 r)+2 r^4}
   \;.
\eeaa

For $r>0$ and   $\Delta_r > 0$ it holds that
\bea
 \displaystyle \frac{ \Delta_r  \Sigma  }{
   (a^2+r^2)^2 } \le \big|g_{tt} - \frac{g_{t\varphi}^2}{g_{\varphi\varphi}}\big|
    \le  \frac{  \Delta_r  \Sigma }{  r \left(a^2 (2 m+r)+r^3\right)}
    \;,
    &
\eeal{23VI12.1}
with the infimum attained at $ \theta\in\{0,\pi\}$ and maximum at $\theta=\pi/2$.

In the region $r>0$, $\Delta_r >0$ consider any vector
$$
 X=X^t\partial_t +X^r\partial_r +X^\theta \partial_\theta + X^\varphi \partial_\varphi
$$
which is causal for the metric $g$. Let $\Omega(r,\theta) $ be any \strictlyx  positive function. Since both $g_{\theta\theta}$ and the first term in \eq{22VI12.4} are  positive, while the coefficient of $dt^2$ in \eq{22VI12.4} is negative, we have
\bean
 0
  & \ge &
   \Omega^2 g(X,X) = \Omega^2 g_{\mu\nu}X^\mu X^\nu
   \ge   \Omega^2\bigg( g_{tt} - \frac{g_{t\varphi}^2}{g_{\varphi\varphi}}\bigg) (X^t)^2 +  \Omega^2 g_{rr} (X^r)^2
\\
 & \ge &
  -\sup_{\theta} \big(\Omega^2\big| g_{tt} - \frac{g_{t\varphi}^2}{g_{\varphi\varphi}}\big|\big) (X^t)^2 +  \inf_{\theta} \left(\Omega^2 g_{rr}\right) (X^r)^2
 \;.
\eeal{22VI12.5}

To guarantee the requirements of the definition of a projection diagram, it is simplest to choose $\Omega$
so that both extrema in \eq{22VI12.5} are attained at the same value of $\theta$, say $\theta_*$,  while keeping those features of the coefficients which are essential for the problem at hand. It is convenient, but not essential, to have $\theta_*$ independent of $r$.
We will make the choice
\bel{23VI12.1om}
\Omega^2 =\frac{r^2+a^2}{\Sigma}
 \;,
\ee
but other choices are possible, and might be more convenient for other purposes. (The $\Sigma$ factor has been included to get rid of the angular dependence in  $\Omega^2 g_{rr}$, while the numerator has been added to ensure that the metric coefficient $\gamma_{rr}$ in \eq{22VI12.6} tends to one as $r$ recedes to infinity, reflecting the asymptotic behaviour for large $r$ of the corresponding function $\hat F$ in \eq{Krakgen2VII12.a2}.)
With this choice of $\Omega$, \eq{22VI12.5} is equivalent to the statement that

\bel{25VI12.3}
 \pi_*(X):=  X^t \partial_t + X^r \partial_r
\ee
is a causal vector in the two-dimensional Lorentzian metric
\bel{22VI12.6}
\gamma:=-    \frac{  \Delta_r  (r^2+a^2) }{  r \left(a^2 (2 m+r)+r^3\right)}   \,  dt^2 + \frac{  (r^2 + a^2) }{ \Delta_r } dr^2
 \;.
\ee
Using the   methods of Walker~\cite{Walker}, as reviewed in Section~\ref{S2VII12.1}, in the region $r_+<r<\infty$,  the metric $\gamma$ is conformal to a flat metric on the interior of a diamond, with the conformal factor extending  smoothly across that part of its  boundary at which $r\to r_+$ when $|a|<m$. This remains true when $|a|=m$ except at the leftmost corner  $i^0_L$ of Figure~\ref{F2dMc2VII12}.

To avoid ambiguities, at this stage $\pi$ is the projection map $(t,r,\theta,\varphi)\mapsto (t,r)$. The fact that $g$-causal curves are mapped to $\gamma$-causal curves follows from the construction of $\gamma$. In order to prove the lifting property, let $ \sigma(s)=(t(s),r(s))$ be a $\gamma$-causal curve, then the curve
$$
  (t(s),r(s),\pi/2,\varphi(s))
  \;,
$$
where
$\varphi(s)$ satisfies
$$
 \frac{d\varphi}{ds} = -  \frac {g_{t\varphi}}{g_{\varphi\varphi}} \frac{ dt}{ds}
$$
is a $g$-causal curve which projects to $\sigma$.

For causal vectors in the region $r>0$, $\Delta_r <0$, we have instead
\bean
 0
  & \ge &
   \Omega^2 g(X,X)
   \ge   \Omega^2\bigg( g_{tt} - \frac{g_{t\varphi}^2}{g_{\varphi\varphi}}\bigg) (X^t)^2 +  \Omega^2 g_{rr} (X^r)^2
\\
 & \ge &
  \inf_{\theta} \left(\Omega^2\bigg| g_{tt} - \frac{g_{t\varphi}^2}{g_{\varphi\varphi}}\bigg|\right) (X^t)^2 -  \sup_{\theta} \left(\Omega^2 |g_{rr}|\right) (X^r)^2
 \;.
\eeal{22VI12.7}
Since the inequalities in \eq{23VI12.1} are reversed when $\Delta_r<0$, choosing the same factor $\Omega$ one concludes again that   $X^t \partial_t + X^r \partial_r$ is   $\gamma$-causal in the metric \eq{22VI12.6} whenever it is in the metric $g$.
Using again~\cite{Walker}, in the region $r_-<r<r_+$, such a metric is conformal to a
a flat two-dimensional metric on the interior of a diamond, with the conformal factor extending smoothly across those parts of its  boundary where $r\to r_+$ or $r\to r_-$.

When $|a|<m$ the metric coefficients in $\gamma$ extend analytically from the $(r>r_+)$--range to the $(r_-<r<r_+)$--range. As described in Section~\ref{S2VII12.1}, one can then smoothly glue together four diamonds as above to a single  diamond on which $r_-<r<\infty$.

The singularity of $\gamma$ at $r=0$ reflects the fact that the metric $g$ is singular at $\Sigma=0$. This singularity persists even if $m=0$, which might at first seem surprising since then there is no geometric singularity at $\Sigma=0$ anymore~\cite{CarterKerr}. However, this singularity of $\gamma$  reflects the singularity of the associated coordinates on Minkowski space-time, with the set $r=0$ in the projection metric corresponding to a boundary of the projection diagram.

\begin{figure}[t]
 \begin{center}
\includegraphics[scale=0.5]{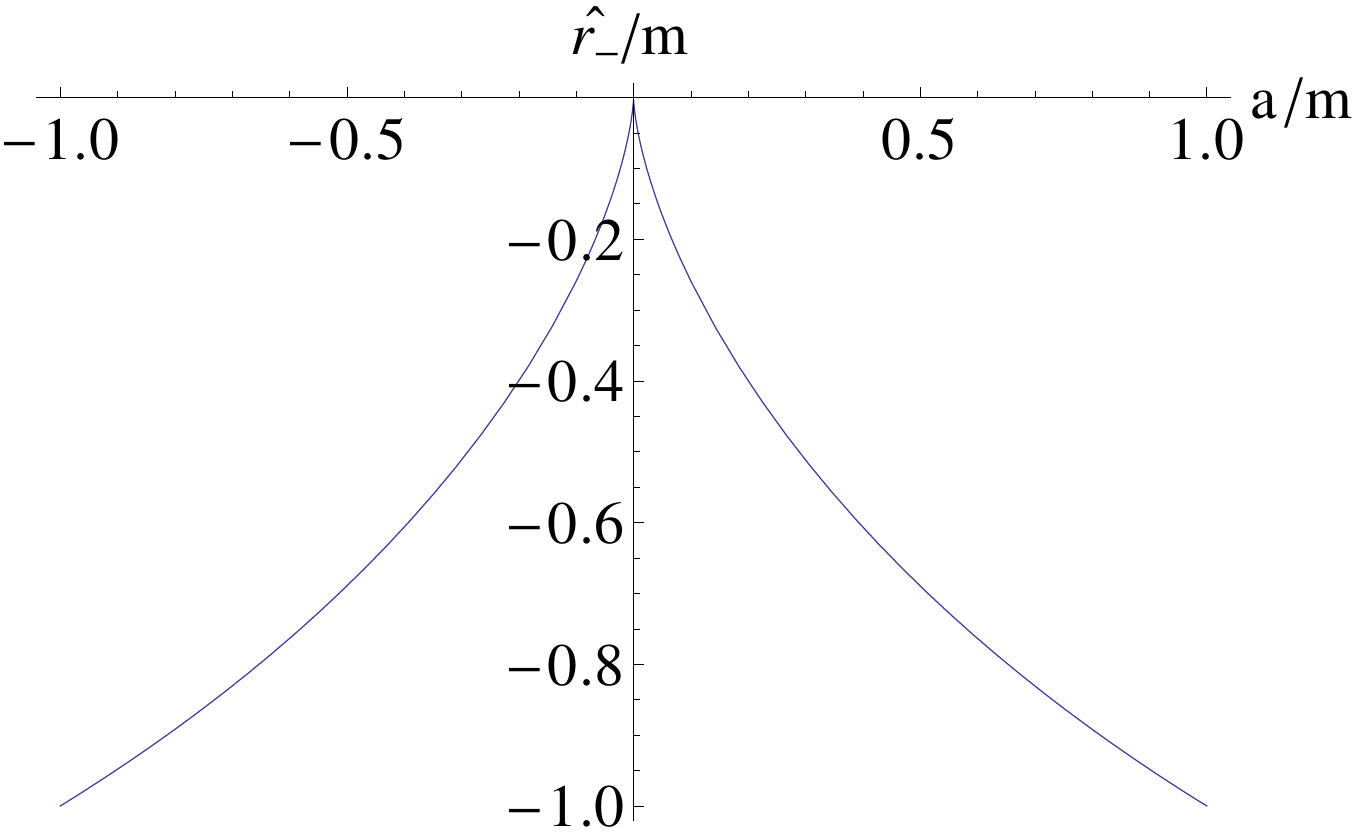}%
\caption{\label{F25VI12.2} The radius of the time-machine ``left boundary" $\hat r_-/m$ as a function of $a/m$.}
\end{center}
\end{figure}

\begin{figure}
\begin{center}
{\includegraphics[scale=.7,angle=90]{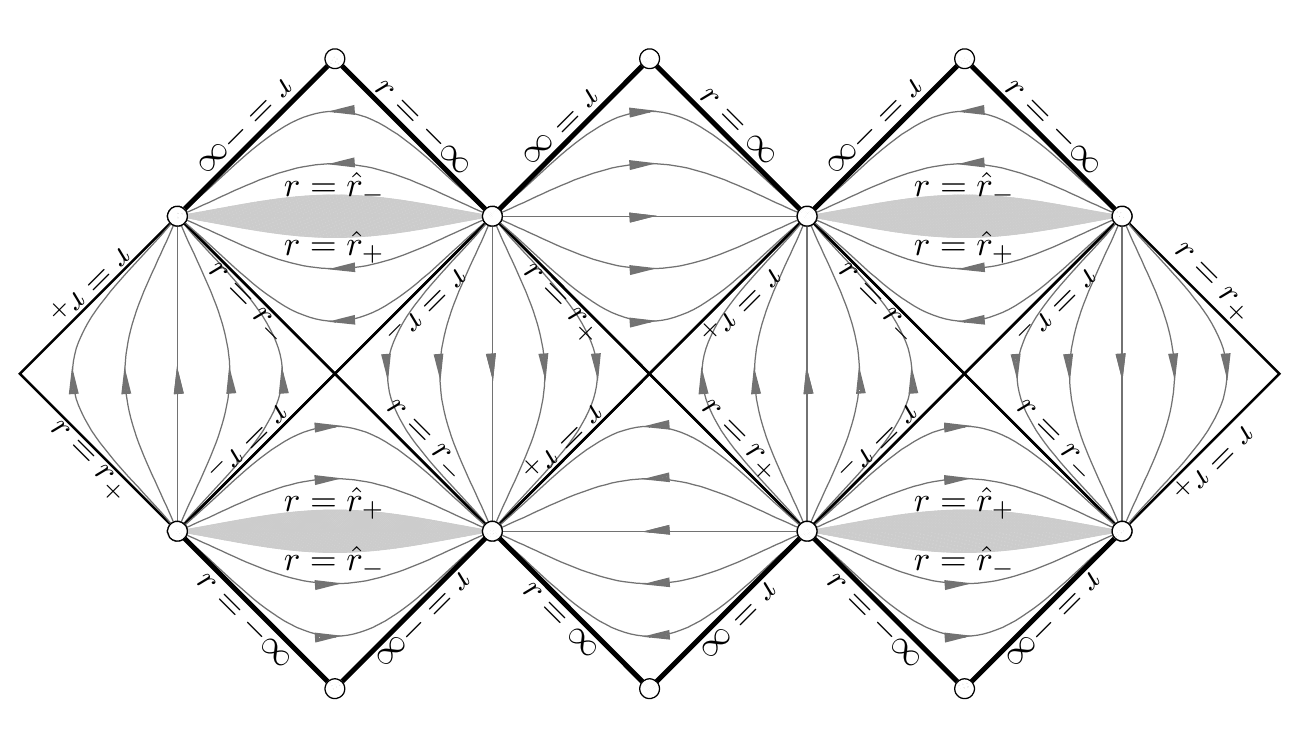}}%
{\includegraphics[scale=.7,angle=90]{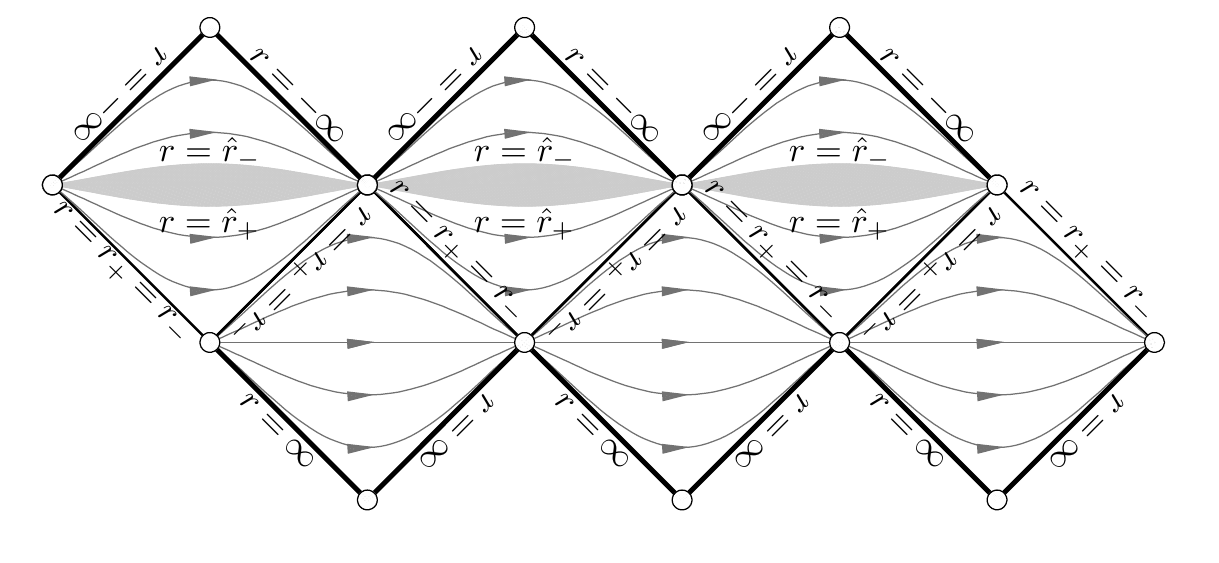}}
\caption{A projection diagram for the Kerr-Newman    metrics with two distinct zeros of $\Delta_r$ (left diagram) and  one double zero  (right diagram); see Remark~\ref{R25VII12.1}. {\color{red} }}
\label{F25VI12.1}
\end{center}
\end{figure}

For $r< 0$ we have $\Delta_r>0$, and the inequality \eq{22VI12.5} still applies in the region where $\partial_\varphi$ is spacelike. Here one needs to keep in mind  the non-empty \emph{Carter time-machine} set (compare~\eq{27VII12.3}),
\bean
\mcV &:= &
 \{g_{\varphi\varphi}< 0 \}
\\
  & = &
    \big\{  r\le 0 \;,\ \cos(2\theta) < -\frac{ a^4+2 a^2 m r+3 a^2 r^2+2 r^4}{a^2 \Delta_r}\;,
    \nonumber
\\
&&  \phantom{xxx} \Sigma \ne 0\;,\  \sin (\theta)\ne 0
 \big\}
 \;,
\eeal{25VI12.4}
on which the Killing vector $\partial_\varphi$ (which has $2\pi$-periodic orbits) is timelike.  The projection of the closure of this region to a two-dimensional diagram should be considered to be  a singular set. But causality is restored regardless of the value of $\theta$ if we remove from $\mcM$  the closure of $\mcV$: Setting
$$
 \mcU:= \mcM\setminus \overline{\mcV}
 \;,
$$
throughout $\mcU$
we have
\bel{25VI12.1}
   \frac{ a^4+2 a^2 m r+3 a^2 r^2+2 r^4}{a^2 \left(a^2-2 m r+r^2\right)}
   > 1
  \quad
  \Longleftrightarrow
  \quad
 {  r \left(a^2 (2 m+r)+r^3\right)} > 0
    \; .
\ee
Equivalently,
\bel{25VI12.2}
r <\hat r_-:= \frac{\sqrt[3]{\sqrt{3} \sqrt{a^6+27 a^4 m^2}-9 a^2 m}}{3^{2/3}}-\frac{a^2}{\sqrt[3]{3} \sqrt[3]{\sqrt{3} \sqrt{a^6+27 a^4
   m^2}-9 a^2 m}} <0
    \;,
\ee
see Figure~\ref{F25VI12.2}.
In the region $r<\hat r_-$ the inequalities \eq{23VI12.1} hold again, and so the projected vector $\pi_*(X)$ as defined by \eq{25VI12.3} is causal, for $g$-causal $X$, in the metric $\gamma$ given by \eq{22VI12.6}. One concludes that  the four-dimensional region $\{-\infty<r<r_-\}$ has the causal structure which projects to those diamonds of, e.g., Figure~\ref{F25VI12.1} which contain a shaded region. Those shaded regions, which  correspond both  to the singularity $r=0$, $\theta=\pi/2$  and to the time-machine region $\mcV$ of \eq{25VI12.4}, belong to $\mcW=\pi(\mcM)$ but \emph{not} to $\pi(\mcU)$.
Causality within the shaded region is \emph{not} represented in any useful way by a flat two-dimensional metric there, as causal curves can exit this region earlier, in Minkowskian time on the diagram, than they entered it. This  results in causality violations throughout the enclosing diamond unless the shaded region is removed.

The projection diagrams for the usual maximal extensions of the Kerr-Newman metrics can be found in Figure~\ref{F25VI12.1}.

\begin{Remark}
 \label{R25VII12.1}
 {\rm
 Let us make some general remarks concerning  projection diagrams   for the Kerr-Newman family of metrics, possibly with a non-vanishing cosmological constant $\Lambda$.
The shaded regions in figures such  as Figure~\ref{F25VI12.1} and others contain the singularity $\Sigma=0$ and the time-machine set $\{g_{\varphi\varphi}<0\}$, they belong to the set $\mcW=\pi(\mcM)$ but do  \emph{not} belong to the set $\pi(\mcU)$, on which causality properties of two-dimensional Minkowski space-time reflect those of $\mcU\subset \mcM$.  We emphasise that there are closed timelike curves through every point in the preimage under $\pi$ of the entire diamonds containing the shaded areas. On the other hand, if the preimages of the shaded region are removed from $\mcM$, the causality relations in the resulting space-times are accurately represented by the diagrams, which are then proper.

The parameters $\hat r_\pm $ are determined by the mass and the charge parameters (see \eq{26VII12.1}), with $\hat r_+=0$ when the charge $e$ vanishes, and $\hat r_+$  positive otherwise.
 The boundaries $r=\pm \infty$ correspond to smooth conformal boundaries at infinity, with causal character determined by $\Lambda$. The arrows indicate  the spatial or timelike character of the orbits of the isometry group.

Maximal diagrams are obtained when continuing the diagrams shown in all allowed directions. It should be kept in mind that the resulting subsets of $\R^2$ are not simply connected in some cases, which implies that many alternative non-isometric maximal extensions of the space-time can be obtained by taking various coverings of the planar diagram. One can also make use of the symmetries of the diagram to produce distinct quotients.
\qed
}
\end{Remark}
\subsubsection{Conformal diagrams for a class of two-dimensional submanifolds of Kerr space-time}

One can find e.g.\ in~\cite{CarterlesHouches,HE} conformal diagrams for the symmetry axes in the maximally extended Kerr space-time.
These diagrams are identical with those of Figure~\ref{F25VI12.1}, except for the absence of shading. (The authors of~\cite{CarterlesHouches,HE} seem to indicate that the subset $r=0$  plays a special role in their diagrams, which is not the case as the singularity $r=\cos\theta=0$ does not intersect the symmetry axes.) Now, the symmetry axes are  totally geodesic submanifolds, being the collection of fixed points of the isometry group generated by the rotational Killing vector field. They can be thought of as the submanifolds  $\theta=0$ and $\theta=\pi$  (with the remaining angular coordinate irrelevant then)  of the extended Kerr space-time.
As such, another totally geodesic two-dimensional submanifold in Kerr is the equatorial plane $\theta=\pi/2$, which is the set of fixed points of the isometry $\theta\mapsto \pi-\theta$. This leads one to enquire
  about the global structure of this submanifold or, more generally, of various families of two-dimensional submanifolds on which $\theta$ is kept fixed. The discussion that follows appears to have some interest of its own. More importantly for us, it illustrates clearly the distinction between projection diagrams, in which one projects-out the $\theta$ and $\varphi$ variables, and   conformal diagrams for submanifolds where $\theta$, and $\varphi$ or the angular variable $\tilde \varphi$ of \eq{16VIII12.1x} below, are fixed.

An obvious family of two-dimensional Lorentzian submanifolds to consider is that of  submanifolds, which we denote as $N_{\theta, \varphi }$,  which are obtained by keeping $\theta$ and $\varphi$ fixed. The metric, say $g(\theta)$, induced by the Kerr metric on  $N_{\theta,\varphi}$ reads
\begin{eqnarray}
  g(\theta)  &=& - \frac{\Delta_r -a^2 \sin ^2(\theta )}{\Sigma }dt^2  +\frac{\Sigma }{\Delta_r } dr^2  =: -F_1(r)dt^2 + \frac {dr^2} {F_2(r)}
  \;.
  \label{31X12.1}
\end{eqnarray}
For $m^2-a^2 \cos^2(\theta)>0$ the function $F_1$ has two first-order zeros at the intersection of $N_{\theta, \varphi }$  with the boundary of the ergoregion $\{g(\partial_t,\partial_t)> 0\}$:
\beq
  \label{31X12.4}
    {r}_{\theta,\pm} = m \pm \sqrt{m^2-a^2 \cos^2(\theta)}
    \;.
\eeq
The key point is that these zeros are distinct from those of $F_2$ if $\cos^2\theta\ne 1$, which we assume in the remainder of this section. Since ${r}_{\theta,+}$ is larger than the largest zero of $F_2$,
the metric $g(\theta)$ is a priori only defined for $r>{r}_{\theta,+}$. One checks that its Ricci scalar diverges as $(r-{r}_{\theta,+})^{-2}$ when ${r}_{\theta,+}$ is approached,
therefore those submanifolds do not extend smoothly across the ergosphere, and will thus be of no further interest to us.

We consider, next, the two-dimensional submanifolds, say
$\tilde N_{\theta,\tilde \varphi}$,    of the Kerr space-time obtained by keeping $\theta $ and $\tilde{\varphi} $ fixed, where $\tilde \varphi$ is a new angular coordinate defined as
\bel{16VIII12.1x}
    d\tilde \varphi = d\varphi +  \frac {a  }{\Delta_r}\, dr\,.
\ee
Using further the coordinate $v$ defined as
\bea
  \label{31X12.11}
  dv &=& dt + \frac{(a^2+r^2)}{\Delta_r} dr \;,
\eea
the metric, say $\tilde g(\theta)$, induced on $\tilde N_{\theta,\tilde \varphi}$ takes the form
\bean
    \tilde g(\theta)
    \nonumber
    &=& - \frac{\tilde F(r) }{\Sigma} dv^2  + 2 dv dr
    \\
    &=& -\frac{\tilde F(r)}{\Sigma} dv\left(dv-2 \frac{\Sigma}{\tilde F(r)}dr\right)
    \;,
  \label{g2}
\eea
where $\tilde F(r):=r^2 + a^2 \cos^2(\theta) -  2 m r$.  The zeros of $\tilde F(r)$ are again given by \eq{31X12.4}.
Setting
\bea
\label{trafo2}
  du = dv-2 \frac{\Sigma}{\tilde F(r)}dr
\eea
brings \eq{g2} to the form
$$
  \tilde g(\theta)= -\frac{\tilde F(r)}{\Sigma} dv du
   \;.
$$
The usual Kruskal-Szekeres type of analysis applies to this metric, leading to a conformal diagram as in the left Figure~\ref{F25VI12.1} \emph{with no shadings}, and with $r_\pm$ there replaced by ${r}_{\theta,\pm}$, \emph{as long as $\tilde F$ has two distinct zeros}.

Several comments are in order:

First, the event horizons \emph{within $\tilde N_{\theta,\tilde \varphi}$} \emph{do not} coincide with the intersection of the event horizons of the Kerr space-time with $\tilde N_{\theta,\tilde \varphi}$. This is not difficult to understand by noting that the class of causal curves that lie within $\tilde N_{\theta,\tilde \varphi}$ is smaller than the class of causal curves in space-time, and there is therefore no a priori reason to expect that the associated horizons will be the same. In fact, is should be clear that the event horizons within $\tilde N_{\theta,\tilde \varphi}$ should be located on the boundary of the ergoregion, since in two space-time dimensions the boundary of an ergoregion is necessarily a null hypersurface. This illustrates the fact that conformal diagrams for submanifolds might fail to represent correctly the location of horizons. The reason that the conformal diagrams for the symmetry axes correctly reflect the global structure of the space-time is an accident related to the fact that the ergosphere touches the event horizon there.

This last issue acquires a dramatic dimension for
extreme Kerr black holes, for which $|a|=m$, where for $\theta\in (0,\pi)$ the global structure of maximally extended $\tilde N_{\theta,\tilde \varphi}$'s is represented by   an unshaded version of the left  Figure~\ref{F25VI12.1}, while the conformal diagrams for the axisymmetry axes are given by the
unshaded version of the right  Figure~\ref{F25VI12.1}.

\begin{figure}
\begin{center}
{\includegraphics[scale=.7,angle=0]{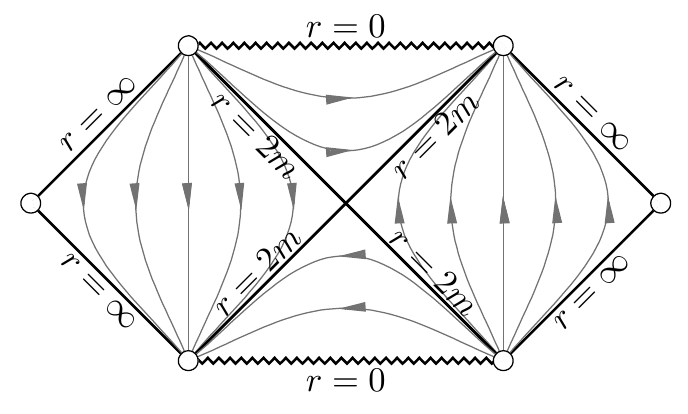}}%
{\includegraphics[scale=.7,angle=0]{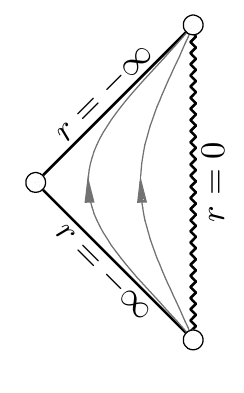}}
\caption{The conformal diagram for a maximal analytic extension of the metric  induced by the Kerr metric, with arbitrary $a\in \R$, on the submanifolds of constant angle $\tilde \varphi$ within the equatorial plane $\theta=\pi/2$, with $r>0$ (left) and $r<0$ (right).
{\color{red}
}}
\label{F25VI12.12}
\end{center}
\end{figure}

Next, another dramatic change arises in the global structure of the $\tilde N_{\theta,\tilde \varphi}$'s with $\theta=\pi/2$. Indeed, in this case we have ${r}_{\theta,+}=2m$, as in Schwarzschild space-time,  and ${r}_{\theta,-}=0$, \emph{regardless of whether the metric is underspinning, extreme, or overspinning}. Since ${r}_{\theta,-}$ coincides now with the location of the singularity, $\tilde N_{\theta,\tilde \varphi}$ acquires two connected components, one where $r>0$ and a second one with $r<0$. The conformal diagram of the first one is identical to that of the Schwarzschild space-time with positive mass, while the second is identical to that of Schwarzschild with negative mass, see Figure~\ref{F25VI12.12}.
We thus obtain the unexpected conclusion, that \emph{the singularity $r=\cos(\theta)=0$ has a spacelike character when approached with positive $r$ within the equatorial plane, and a timelike one when approached with negative $r$ within that plane}. This is rather obvious in retrospect, since the metric induced by Kerr on  $\tilde N_{\pi/2,\tilde \varphi}$ coincides, when $m>0$, with the one induced by the Schwarzschild metric with positive mass in the region $r>0$ and with the Schwarzschild metric with negative mass $-m$ in the region $r<0$.

\begin{figure}
\begin{center}
{\includegraphics[scale=.7,angle=0]{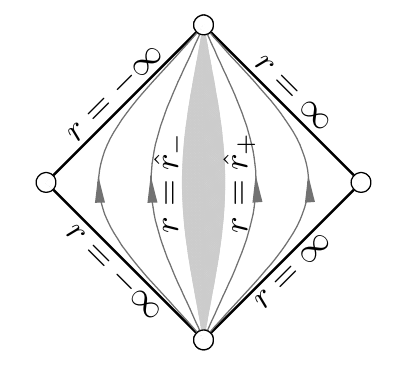}}%
\caption{A projection diagram for overspinning Kerr-Newman space-times.
{\color{red}}}
\label{F2XI12.1}
\end{center}
\end{figure}

Note finally that, surprisingly enough, even for overspinning Kerr metrics  there will be a range of angles $\theta$ near $\pi/2$ so that $\tilde F$ will have two distinct first-order zeros. This implies that, for such $\theta$, the global structure of maximally extended $ \tilde N_{\theta, \tilde\varphi}$'s will be similar to that of the corresponding submanifolds of the underspinning Kerr solutions. This should be compared with the projection diagram for overspinning Kerr space-times, to be found in Figure~\ref{F2XI12.1}.

 \subsubsection{The orbit space-metric on $\mcM/\Uone$}

Let $h$ denote the tensor field obtained by quotienting-out in the Kerr metric $g$ the $\eta:=\partial_\varphi$ direction,
\bel{22VI12.1}
 h(X,Y) = g(X,Y) - \frac {g( X,\eta) g(Y,\eta)}{g(\eta,\eta)}
 \;.
\ee
The tensor field $h$ projects to the natural quotient metric on the manifold part of $\mcM/\Uone$.
In the region where $\eta$ is spacelike, the quotient space $\mcM/\Uone$ has the natural structure of a manifold with boundary, where the boundary is the image, under the quotient map, of the axis of rotation %
$$
 \mcA:= \{\eta = 0\}
 \;.
$$
Using $t,r,\theta$ as coordinates on the quotient space we find a diagonal metric
\bel{22VI12.3}
 h = h_{tt} dt^2   +\frac{\Sigma}{\Delta_r}dr^{2}
+\Sigma d\theta ^{2}\;,
\ee
where
$$
 h_{ tt} = g_{tt} - \frac{g_{t\varphi}^2}{g_{\varphi\varphi}}
   \;,
$$
as in \eq{22VI12.4}.
Thus, the metric $\gamma$ of Section~\ref{S2VII12.2} is directly constructed out of the $(t,r)$--part of the quotient-space metric $h$. However, the analogy is probably misleading as there does not seem to be any direct correspondence between the quotient space $\mcM/\Uone$ and the natural manifold as constructed in Section~\ref{S2VII12.2} using the metric $\gamma$.%
\footnote{Once this work was written it was pointed out to us that the idea of using the Penrose diagram for the quotient-space metric has been used in~\cite{GibbonsHerdeiro}. The Penrose-Carter conformal diagram of Section~4.6 of \cite{GibbonsHerdeiro} coincides with a projection diagram for the BMPV metric, but our interpretation of this diagram differs.}

\subsection{The Kerr-Newman metrics}
 \label{S2VII12.4}

The analysis of the Kerr-Newman metrics is essentially identical: The metric takes the same general form \eq{Kerr2x},
 except that now
$$
 \Delta_r = r^2+a^2+e^2-2 m r=:(r-r_+)(r-r_-)
  \;,
$$
and we assume that $e^2 + a^2 \le m$ so that the roots are real.   We have
\begin{eqnarray}
  &
   \displaystyle
  g_{\varphi\varphi}
     =\frac{\sin ^2(\theta ) \left(\left(r^2+a^2\right)^2-a^2 \Delta_r  \sin ^2(\theta )\right)}{\Sigma }
    \;,
    &
 \label{19VII12.1}
\\
&
   \displaystyle
 g_{tt}-\frac{g_{t\varphi}^2}{g_{\varphi \varphi}}
 =
 -\frac{\Delta_r  \Sigma }{\left(r^2+a^2\right)^2-a^2 \Delta_r  \sin ^2(\theta )}
 \;,
 &
  \label{19VII12.2}
\eea
and note that the sign of the denominator in \eq{19VII12.2} coincides with the sign of $g_{\varphi\varphi}$.
Hence
$$
\mbox{\rm sign}(g_{tt}-\frac{g_{t\varphi}^2}{g_{\varphi \varphi}})=-
\mbox{\rm sign}(\Delta_r)
\mbox{\rm sign}(g_{\varphi\varphi})
 \;.
$$
For $g_{\varphi \varphi} > 0$, which is the main region of interest, we conclude that the minimum of $(g_{tt}-\frac{g_{t\varphi}^2}{g_{\varphi \varphi}})\Sigma^{-1}\Delta_r^{-1}$ is assumed at $\theta = \frac{\pi}{2}$ and the maximum at $\theta = 0,\pi$, so for all $r$ for which $g_{\varphi \varphi} > 0$  we have

\beq
-\frac{\Delta_r  \Sigma }{\left(r^2+a^2\right)^2-a^2 \Delta_r } \leq g_{tt}-\frac{g_{t\varphi}^2}{g_{\varphi \varphi}} \leq -\frac{\Delta_r  \Sigma }{\left(r^2+a^2\right)^2} \;.
\eeq
Choosing the conformal factor as
$$
\Omega^2 =\frac{r^2 + a^2 }{\Sigma}
$$
we obtain, for $g$-causal vectors $X$,
\bean
 \nonumber
0
  & \geq &
   \Omega^2 g(X,X) = \Omega^2 g_{\mu\nu}X^\mu X^\nu
   \geq   \Omega^2\bigg( g_{tt} - \frac{g_{t\varphi}^2}{g_{\varphi\varphi}}\bigg) (X^t)^2 +  \Omega^2 g_{rr} (X^r)^2
\\
 & \geq & -\frac{\Delta_r \left(r^2+a^2\right) }{\left(r^2+a^2\right)^2-a^2 \Delta_r } (X^t)^2 + \frac{\left(r^2+a^2\right)}{\Delta_r}(X^r)^2
 \;.
\eea
This leads to the following projection metric
\bean
\gamma &:=& - \frac{  \Delta_r \left(r^2+a^2\right) }{\left(r^2+a^2\right)^2-a^2 \Delta_r }  \,  dt^2 + \frac{ \left(r^2+a^2\right) }{ \Delta_r } dr^2 \\
&=& - \frac{  \Delta_r  \left(r^2+a^2\right)}{a^2 \left(r (2 m+r)-e^2\right)+r^4}  \,  dt^2 + \frac{ \left(r^2+a^2\right)  }{ \Delta_r } dr^2
 \;,
\eea
which is Lorentzian if and only if $r$ is such that $g_{\varphi\varphi}> 0 $ for all $\theta \in [0,\pi]$.
Now, it follows from \eq{19VII12.1} that
$g_{\varphi\varphi}$ will have the \emph{wrong} sign if
\begin{eqnarray}
\label{18VII12.1}
  0
    &>&  \left(r^2+a^2\right)^2-a^2 \Delta_r  \sin ^2(\theta )
     \;.
\end{eqnarray}
This does not happen when $\Delta_r\le0$, and hence in a neighborhood of both horizons. On the other hand, for $\Delta_r>0$, a necessary condition for \eq{18VII12.1} is
\begin{eqnarray}
\label{18VII12.1a}
  0
    &>&  \left(r^2+a^2\right)^2-a^2 \Delta_r  = {r}^{4}+{r}^{2}{a}^{2}+2\,mr{a}^{2}-{a}^{2}{e}^{2} =:f(r)
     \;.
\end{eqnarray}
The second derivative of $f$ is \strictlyx  positive, hence $f'$ has exactly one real zero.  Note that $f$ is  strictly   smaller than the corresponding function for the Kerr metric, where $e=0$, thus the interval where $f$ is \strictlyx  negative encloses the corresponding interval for Kerr. We conclude that $f$ is negative on an interval $(\hat r_-,\hat r_+)$, with $\hat r_-<0<\hat r_+<r_-$.

The corresponding projection diagrams are identical to those of the Kerr space-time, see Figure~\ref{F25VI12.1}, with the minor modification that the region to be excised from the diagram is  $\{r\in (\hat r_-,\hat r_+)\}$, with now $\hat r_+>0$, while we had $\hat r_+=0$ in the uncharged case.

\subsection{The Kerr - de Sitter metrics}
 \label{S2VII12.5}

The Kerr - de Sitter metric in Boyer-Lindquist-like coordinates reads~\cite{Demianski,Carterseparable}
\begin{eqnarray}
  \label{kds}
  \nonumber
    g &=& \frac{\Sigma}{\Delta_r} dr^2+\frac{\Sigma}{\Delta_\theta} d\theta^2
      +\frac{\sin ^2(\theta)}{\Xi^{2} \Sigma }\Delta_\theta \left(a dt-(r^2+a^2)d\varphi\right)^2  \\
    &\quad& - \frac{1}{\Xi^{2} \Sigma }\Delta_r \left(dt-a \sin ^2(\theta) \, d\varphi\right)^2
     \;,
\end{eqnarray}
where
\begin{eqnarray}
\label{kds2}
    \Sigma = r^2+a^2\cos ^2(\theta) \vphantom{\frac11}\;, \qquad \Delta_r = (r^2+a^2)\left(1-\frac{\Lambda}3 r^2\right)-2 \mu \Xi r \;,
\end{eqnarray}
and
\begin{eqnarray}
 \label{kds3}
    \Delta_\theta = 1+\frac{\Lambda}3 a^2 \cos ^2(\theta) \;, \qquad \Xi = 1+\frac{\Lambda}3 a^2 \;,
\end{eqnarray}
for some real parameters $a$ and $\mu$, where $\Lambda $ is the cosmological constant.
In this section we assume $\Lambda>0$ and $a\ne 0$. By a redefinition  $\varphi\mapsto-\varphi$ we can always achieve $a>0$, similarly changing $r$ to $-r$ if necessary we can assume that  $\mu \ge 0$. The case $\mu=0$ leads to the de Sitter metric in unusual coordinates (see, e.g., \cite[Equation~(17)]{AMatzner}).
The inequalities $a>0$ and $\mu > 0$  will be assumed from now on.

The Lorentzian character of the metric should be clear from \eq{kds}; alternatively, one can calculate the determinant of $g$:
\begin{equation}
  \label{det}
  \det(g) = - \frac{\Sigma^2}{\Xi^4}\sin^2{\theta}
  \;.
\end{equation}

We have
\bel{22VII12.6}
 g^{tt} = \frac{g_{rr}\, g_{\theta\theta} \,g_{\varphi\varphi}}{\det (g)}
  = - \frac{ \Xi^4 } {\Delta_\theta } \times \frac 1 {\Delta_r } \times \frac{g_{\varphi\varphi}}{\sin^2{\theta}}
  \;,
\ee
which shows that either $t$ or its negative is a time function whenever $\Delta_r $ and $ {g_{\varphi\varphi}}/{\sin^2{\theta}}$ are \strictlyx  positive. (Incidentally, chronology is violated on the set where $g_{\varphi\varphi}<0$, we will return to this shortly.)  One also has
\bel{22VII12.7}
 g^{rr} = \frac{\Delta_r}{\Sigma}
 \;,
\ee
which shows that $r$ or its negative  is a time function in the region where $\Delta_r<0$.

The character of the principal orbits of the isometry group $\mathbb{R}\times U(1)$ is determined by the sign of the determinant
\begin{equation}
  \det
    \left(
      \begin{array} {cc}
         g_{tt} & g_{\varphi t} \\
         g_{\varphi t} & g_{\varphi \varphi}
      \end{array}
    \right)
  = -\frac{\Delta_r \Delta_\theta}{\Xi^4} \sin^2{\theta} \;.
\end{equation}
Therefore,  for $\sin(\theta) \neq 0$ the orbits are two-dimensional,
 timelike in the regions where $\Delta_r > 0$, spacelike where $\Delta_r<0$, and null where $\Delta_r=0$ once the space-time has been appropriately extended to include the last set.

When $\mu \ne 0$ the set $\{\Sigma=0\}$ corresponds to a geometric singularity in the metric.
 To see this, note that
\bel{22VII12.2}
  g(\partial_t,\partial_t)= \frac{a^2 \sin^2 \theta \Delta_\theta - \Delta_r}{\Sigma\, \Xi^2 }
  = 2\,{\frac {\mu\,r}{\Sigma\,\Xi}} +O(1)
  \;,
\ee
where $O(1)$ denotes  a function which is bounded near $\Sigma=0$.
It follows that for $\mu\ne 0$ the norm of the Killing vector $\partial_t$ blows up as the set $\{\Sigma=0\}$ is approached along the plane $\cos (\theta)=0$, which would be impossible if the metric could be continued across this set in a $C^2$ manner.

The function $\Delta_r$ has exactly two  distinct first-order real zeros when
\bel{25VI12.11x}
\mu^2>\frac{2}{3^5\Xi^2\Lambda}\left(3-a^2\Lambda\right)^3
         \;.
\ee
It has at least two, and up to four, possibly but not necessarily distinct, real roots when
\bel{25VI12.11} a^2 \Lambda \leq 3 \;,
\quad
\mu^2\le\frac{2}{3^5\Xi^2\Lambda}\left(3-a^2\Lambda\right)^3
         \;.
\ee
The negative root $r_1$ is always simple  and \strictlyx negative, the remaining ones are \strictlyx  positive.
We can thus order the roots as
\bel{25VI12.12}
r_1 < 0 < r_2\le r_3 \le r_4
         \;,
\ee
when there are four real ones, and we set $r_3\equiv r_4:=r_2$ when there are only two real roots $r_1<r_2$.
The function $\Delta_r$ is  \strictlyx positive for $r\in (r_1,r_2)$, and for $r\in (r_3,r_4)$ whenever the last interval is not empty; $\Delta_r$ is negative or vanishing otherwise.

It holds that
\begin{eqnarray}
 \label{22VII12.5}
  g_{\varphi \varphi} &= &  \frac{\sin ^2(\theta ) \left(\Delta_\theta\left(r^2+a^2\right)^2-a^2 \Delta_r  \sin ^2(\theta )\right)}{\Xi^2 \Sigma }
\\
   & = &
    \frac{\sin ^2(\theta)}{\Xi} \left( \frac{2a ^2 \mu r \sin ^2(\theta)}{a^2 \cos ^2(\theta)+r^2} +  a^2+r^2 \right)
   \;.
 \label{22VII12.5a}
\end{eqnarray}
The second line is manifestly non-negative for $r\geq 0$, and positive there away from the axis $\sin (\theta)=0$. The first line is manifestly non-negative for $\Delta_r\le0$, and hence also in a neighborhood of this set.

Next
\begin{eqnarray}
   g_{tt} - \frac{g_{t\varphi}^2}{g_{\varphi\varphi}}
    \nonumber
     &=&
       - \frac{\Delta_\theta \Delta_r \Sigma  }{\Xi^2 \left(\Delta_\theta\left(r^2+a^2\right)^2-\Delta_r a^2\sin ^2(\theta)\right)} \\
     &=&
       -\frac{ \Delta_\theta \Delta_r \Sigma  }{\Xi^2\left(A(r)  + B(r) \cos (2 \theta ) \right)}
       \;,
\label{25VI12.7}
\end{eqnarray}
with
\begin{eqnarray}
  A(r) & =  &
 \frac{\Xi }{2}   \left(a^4+3 a^2 r^2+2 r^4
 +2 a^2 \mu r\right)
\;,
\\
 B(r) & = &
 \frac{a^2 }{2} \Xi \left( a^2+r^2 - 2 \mu r  \right)
   \;.
\end{eqnarray}
We have
\begin{eqnarray}
\nonumber
  A(r)+
 B(r) & = &  \Xi  \left(a^2+r^2\right)^2\;,
 \\
  A(r)-
 B(r)
 & = &  {r^2}\Xi \left(a^2+r^2+2 \frac {a^2\mu} r\right)
 \;,
\end{eqnarray}
which confirms that for $r>0$, or for large negative $r$, we have $A > |B| >0 $, as needed for $g_{\varphi\varphi}\ge 0$.
The function
$$
 f(r,\theta):= \frac{ \left(A(r)  + B(r) \cos (2 \theta ) \right)}{ \Delta_\theta  } \equiv \frac{ \left(A(r)  + B(r) \cos (2 \theta ) \right)}{ 1+\frac{\Lambda}3 a^2 \cos ^2(\theta)  }
$$
satisfies
\bel{24VII12.1}
 \frac{\partial f}{\partial \theta} = -  \frac{ a^2 \Xi }{\Delta_\theta^2}\Delta_r \sin(2\theta)
 \;,
\ee
which has the same sign as $-\Delta_r \sin (2\theta)$. In any case, its extrema are achieved at $\theta=0,\,\pi/2$ and $\pi$. Accordingly, this is where the extrema of the right-hand side of \eq{25VI12.7} are achieved as well.
In particular, for  $\Delta_r>0$, we find
\begin{eqnarray}
        \frac{   \Delta_r \Sigma  }{    \left(a^2+r^2\right)^2 }
         \le \Xi^2\left| g_{tt} - \frac{g_{t\varphi}^2}{g_{\varphi\varphi}}\right|
   \le
\frac{\Sigma \Delta_r}{\Xi  r \left(a^2 (2 \mu +r)+r^3\right)}
       \;,
        \nonumber
\\
&&
\label{25VI12.8}
\end{eqnarray}
with the minimum attained at $\theta=0$ and the maximum attained at $\theta=\pi/2$.

To obtain the projection diagram, we can now repeat word for word the analysis carried out for the Kerr metrics on the set $\{g_{\varphi\varphi} >0\}$. Choosing a conformal factor $\Omega^2$ equal to
\bel{25VI2.14}
 \Omega^2 =   \frac {r^2+a^2}\Sigma
\;,
\ee
one is led to a projection metric
\bel{25VI12.9}
\gamma:=-\frac{( {r^2+a^2})\Delta_r}{\Xi^3  r \left(a^2 (2 \mu +r)+r^3\right)}
  \,  dt^2 + \frac{ {r^2+a^2}}{ \Delta_r } dr^2
 \;.
\ee

It remains to understand the set
\begin{eqnarray}
 \nonumber
\mcV &:= &
 \{g_{\varphi\varphi}< 0\}
  \label{22VII12.9x}
\end{eqnarray}
where $g_{\varphi\varphi}$ is negative. To avoid repetitiveness, we will do it simultaneously both for the charged and the uncharged case, where \eq{22VII12.5} still applies (but not  \eq{22VII12.5a} for $e\ne 0$) with $\Delta_r$ given by \eq{26VII12.1}; the Kerr - de Sitter case is obtained by setting $e=0$ in what follows.
A calculation shows that $g_{\varphi\varphi}$ is the product of a non-negative function with
$$
\chi:=2\,{a}^{2}\mu\,r- a^2 e^2 +{r}^{2}{a}^{2}+{r}^{4}+ \left( {r}^{2}{a}^{2}-2\,{a}
^{2}\mu\,r+a^2 e^2+{a}^{4} \right) \cos^2(\theta)
\;.
$$
 This is clearly  \strictlyx  positive for all $r$ and all $\theta\ne \pi/2$ when $\mu =e=0$, which shows that $\mcV=\emptyset$ in this case.

 Next,  the function $\chi$ is sandwiched between the two following functions of $r$, obtained by setting $\cos(\theta)=0$ or $\cos^2(\theta)=1$ in $\chi$:
\beaa
\chi_0
 &:= &
   {r}^{4}+{r}^{2}{a}^
 {2}+2\,{a}^{2}\mu\,r-{a}^{2}{e}^{2}
 \;,
\\
 \chi_1
  & :=  &
   \left( {r}^{2}+{a}^{2}
 \right) ^{2}
 \;.
\eeaa
Hence, $\chi$ is \strictlyx  positive for all $r$ when $\cos^2 (\theta)=1$.
Next, for $\mu>0$ the function  $\chi_0$ is negative for negative $r$ near zero. Further, $\chi_0$ is convex. We conclude that, for  $\mu>0$,  the set on which $
\chi_0 $ is non-positive is a non-empty interval $[\hat r_-,\hat r_+]$ containing the origin.  We have already seen that $g_{\varphi\varphi}$ is non-negative wherever $\Delta_r\le0$, and since $r_2>0$ we must have
 $$r_1< \hat r_-\le\hat r_+ < r_2
 \;.
 $$
In fact, when $e=0$ the value of $\hat r_-$ is given by \eq{25VI12.2} with $m$ there replaced by $\mu$, with $\hat r_-=0$ if and only if $\mu=0$.

We conclude that if $\mu=e=0$ the time-machine set is empty, while if $|\mu| + e^2>0$ there are always  causality violations ``produced" in the non-empty region $\{\hat r_-\le r\le \hat r_+\}$.

The projection diagrams for the Kerr-Newman - de Sitter family of metrics depend upon the number of zeros of $\Delta_r$, and their nature, and can be found in Figures~\ref{F25VI12.4}-\ref{F25VI12.7}.
\begin{figure}
\begin{center}
{\includegraphics[scale=.7,angle=90]{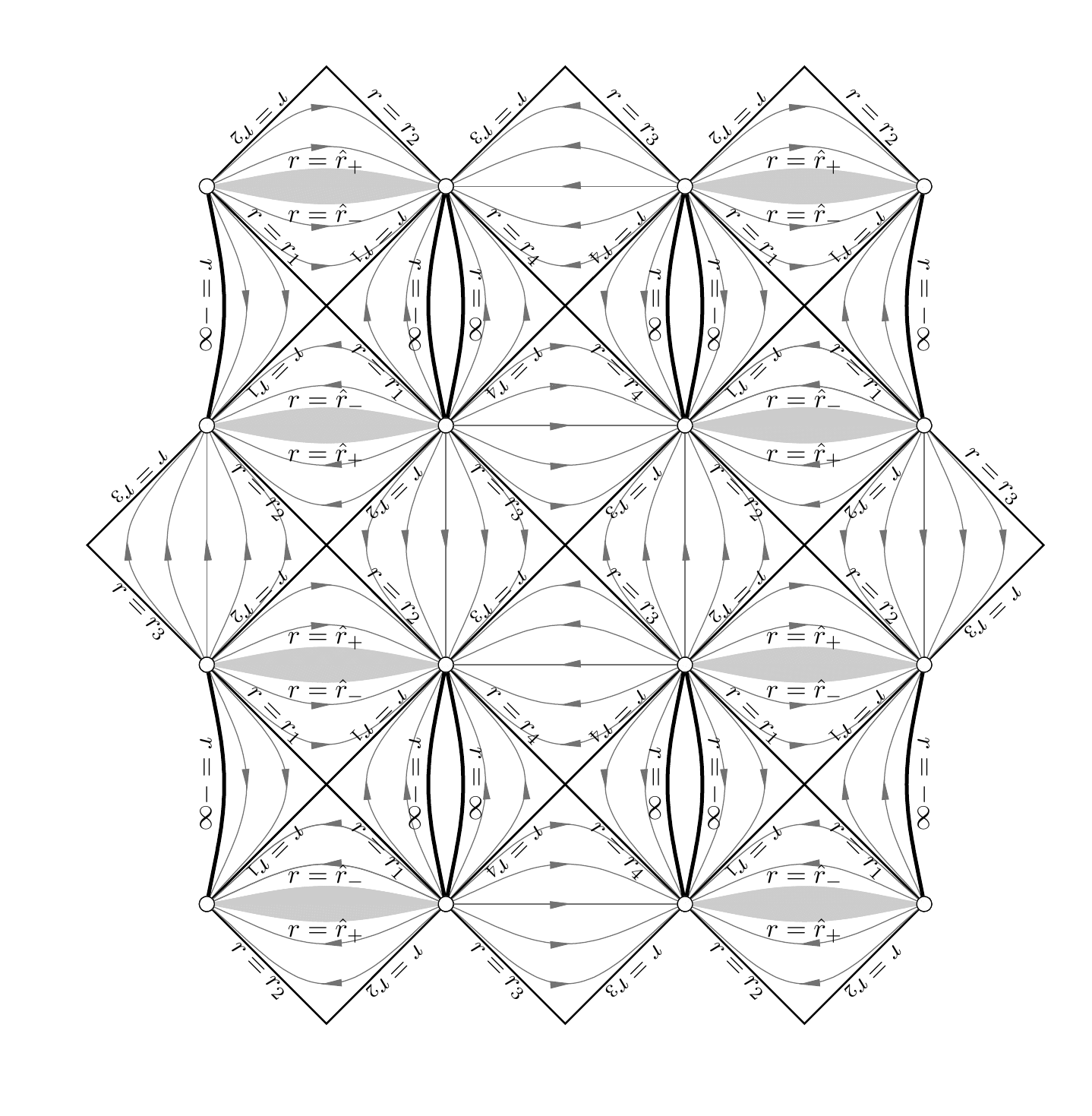}}
\caption{A projection diagram for the Kerr-Newman - de Sitter metric with four distinct zeros of $\Delta_r$; see Remark~\ref{R25VII12.1}. {\color{red} }}
\label{F25VI12.4}
\end{center}
\end{figure}
\begin{figure}
\begin{center}
{\includegraphics[scale=.7,angle=90]{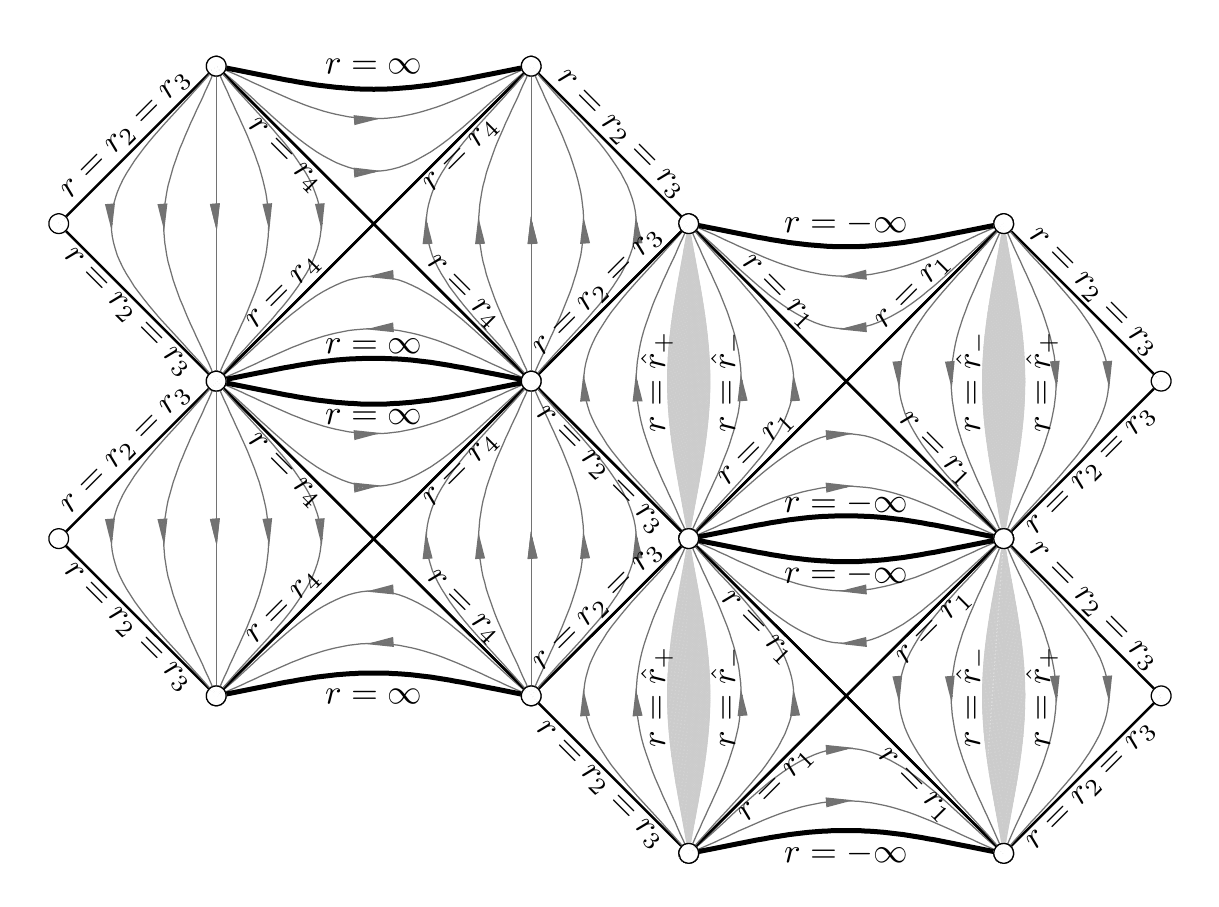}}
\caption{A projection diagram for the Kerr-Newman - de Sitter metrics with three distinct zeros of $\Delta_r$, $r_1<0<r_2=r_3<r_4$; see Remark~\ref{R25VII12.1}. {\color{red}  }}
\label{F25VI12.5}
\end{center}
\end{figure}
\begin{figure}
\begin{center}
{\includegraphics[scale=.7,angle=90]{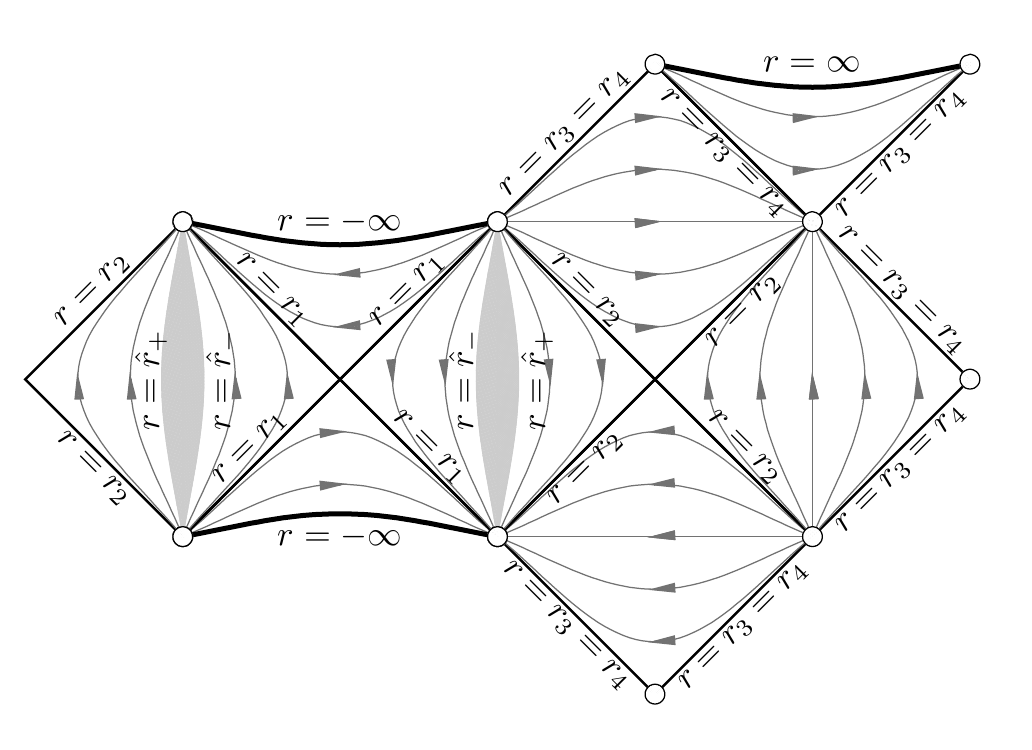}}
\caption{A projection diagram for the Kerr-Newman - de Sitter metrics with three distinct zeros of $\Delta_r$, $r_1<0<r_2<r_3=r_4$; see Remark~\ref{R25VII12.1}.
Note that one cannot continue the diagram simultaneously across all boundaries $r=r_3$ on $\R^2$, but this can be done on an appropriate Riemann surface. {\color{red}}}
\label{F25VI12.6}
\end{center}
\end{figure}
\begin{figure}
\begin{center}
{\includegraphics[scale=.7,angle=0]{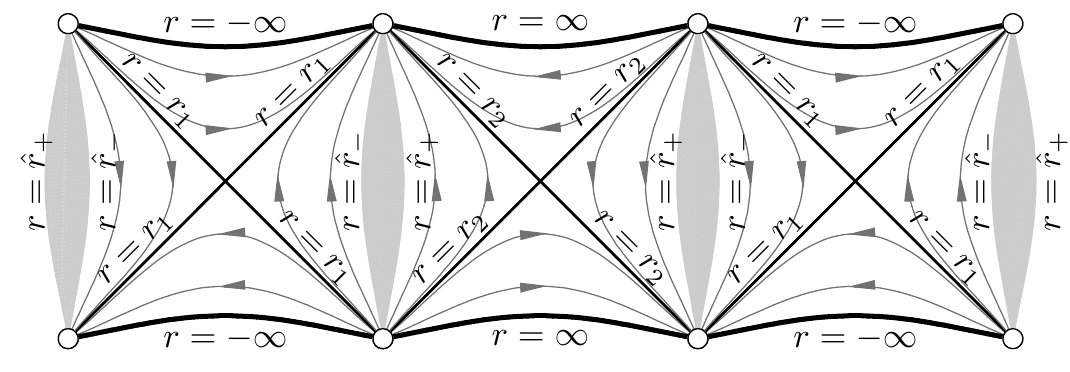}}
\caption{A projection diagram for the Kerr-Newman - de Sitter metrics with two distinct first-order zeros of $\Delta_r$, $r_1<0<r_2 $ and $\mu>0$; see Remark~\ref{R25VII12.1}. The diagram for a first-order zero at $r_1$ and third-order zero at $r_2=r_3=r_4$ would be identical except for the bifurcation surface of the bifurcate Killing horizon at the intersection of the lines $r=r_2$, which does not exist in the third-order case and has therefore to be removed from the diagram. {\color{red}}   }
\label{F25VI12.7}
\end{center}
\end{figure}

\subsection{The Kerr-Newman - de Sitter metrics}
 \label{S2VII12.6}

In the standard Boyer--Lindquist coordinates   the Kerr-Newman -  de Sitter metric  takes the form \eq{kds} \cite{Carterseparable,Exactsolutions2},%
\footnote{The transition  from the formulae in~\cite{Carterseparable} to \eq{kds} is explained in~\cite[p.~102]{CarterlesHouches}.}
 with all the functions as in \eq{kds2}-\eq{kds3} except for  $\Delta_r$, which instead takes the form
\beal{26VII12.1}
  &&\Delta_r =
    \left(1-{\textstyle\frac{1}{3}}\Lambda r^2\right)(r^2+a^2)-2\Xi \mu r+\Xi e^2
    \;,
\eea
where $\sqrt{\Xi}e$ is the electric charge of the space-time. In this section we assume
$$
 \Lambda >0\;,
 \quad
 \mu \ge 0
 \;,
 \quad
 a > 0
 \;,
 \quad
 e\ne 0
 \;.
$$

The calculations of the previous section, and the analysis of zeros of $\Delta_r$, remain identical except for the following equations: First,
\begin{eqnarray}
 \label{22VII12.5N}
  g_{\varphi \varphi} = \frac{\sin ^2(\theta)}{\Xi } \left( \frac{ a ^2 (2  \mu r - e^2)\sin ^2(\theta)}{ a^2 \cos ^2(\theta)+r^2 } +  a^2+r^2 \right)
   \;,
\end{eqnarray}
the sign of which requires further analysis, we will return to this shortly. Next, we still have
\begin{eqnarray}
   g_{tt} - \frac{g_{t\varphi}^2}{g_{\varphi\varphi}}
    \nonumber
     &=&
       - \frac{\Delta_\theta \Delta_r \Sigma  }{\Xi^2 \left(\Delta_\theta\left(r^2+a^2\right)^2-\Delta_r a^2\sin ^2(\theta)\right)} \\
     &=&
       -\frac{ \Delta_\theta \Delta_r \Sigma  }{\Xi^2\left(A(r)  + B(r) \cos (2 \theta ) \right)}
       \;,
\label{25VI12.7N}
\end{eqnarray}
but now
\begin{eqnarray}
  A(r) & =  &
 \frac{\Xi }{2}   \left(a^4+3 a^2 r^2+2 r^4
 +2 a^2 \mu r - a^2 e^2\right)
\;,
\\
 B(r) & = &
 \frac{a^2 }{2} \Xi \left( a^2+r^2 - 2 \mu r  +e^2 \right)
   \;,
\end{eqnarray}
with
\begin{eqnarray}
\nonumber
  A(r)+
 B(r) & = &  \Xi  \left(a^2+r^2\right)^2\;,
 \\
  A(r)-
 B(r)
 & = &  {r^2}\Xi \left(a^2+r^2+2 \frac {a^2\mu} r - \frac{ a^2 e^2}{r^2} \right)
 \;.
\end{eqnarray}
\Eq{24VII12.1} remains unchanged, and for  $\Delta_r>0$, we find
\begin{eqnarray}
        \frac{   \Delta_r \Sigma  }{    \left(a^2+r^2\right)^2 }
         \le \Xi^2\left| g_{tt} - \frac{g_{t\varphi}^2}{g_{\varphi\varphi}}\right|
   \le
\frac{\Sigma \Delta_r}{\Xi   \left(a^2 (2 \mu r - e^2 +r^2)+r^4\right)}
       \;,%
\label{25VI12.8N}
\end{eqnarray}
with the minimum attained at $\theta=0$ and the maximum attained at $\theta=\pi/2$. This leads to the projection metric
\bel{25VI12.9N}
\gamma:=-\frac{\Delta_r}{\Xi^3    \left(a^2 (2 \mu r - e^2 +r^2)+r^4\right)}
  \,  dt^2 + \frac{ 1 }{ \Delta_r } dr^2
 \;.
\ee

We recall that the analysis of the time-machine set $\{g_{\varphi\varphi}<0\}$ has already been carried out at the end of Section~\ref{S2VII12.5},  where it was shown that for  $e\ne 0$  causality violations always exist, and arise from the non-empty region $\{\hat r_-\le r\le  \hat r_+\}$.

The projection diagrams for the Kerr-Newman - de Sitter family of metrics can be found in Figures~\ref{F25VI12.4}-\ref{F25VI12.7}.

\subsection{The Kerr-Newman -  anti de Sitter metrics}
 \label{S2VII12.7}

 We consider the metric \eq{kds}-\eq{kds3}, with however $\Delta_r$ given by \eq{26VII12.1}, assuming that
 $$
  a^2+e^2> 0\;,
   \quad \Lambda < 0
   \;.
 $$
  While the local calculations carried out in Section~\ref{S2VII12.5} remain unchanged, one needs to reexamine
   the occurrence of zeros of $\Delta_r$.

  We start by noting that the requirement that $\Xi\ne 0$ imposes
 $$
  1+ \frac \Lambda 3 a^2 \ne 0
  \;.
 $$
Next, a negative $\Xi$ would lead to a function $\Delta_\theta$ which changes sign. By inspection, one finds that the signature changes from $(-+++)$ to $(+---)$ across these zeros, which implies nonexistence of a coordinate system in which the metric could be smoothly continued there.%
 \footnote{We, and Kayll Lake (private communication), calculated several curvature invariants for the overspinning metrics and found no singularity at $\Delta_\theta=0$. The origin of this surprising fact is not clear to us.}
From now on we thus assume that
\bel{22VII12.1}
 \Xi\equiv  1+ \frac \Lambda 3 a^2 > 0
  \;.
\ee

It is well known that those metrics for which $\Delta_r$ has no zeros are nakedly singular whenever
\bel{29VII12.1}
 e^2 + |\mu| >0
  \;.
\ee
This can, in fact, be easily seen from the following formula for $g_{tt}$ on the equatorial plane:
\bel{28VII12.31}
g_{tt} = {\frac 1 {3{\Xi}^{2}{r}^{2}}} {(-3\,\Xi\,{e}^{2}+6\,\Xi\,\mu\,
   r+ \left( \Lambda\,{a}^{2}-3 \right) {r}^{2}+\Lambda\,{r}^{4}
)}
 \;.
\ee
So, under \eq{29VII12.1}
the norm of the Killing vector $\partial_t$ is unbounded and the metric cannot be $C^2$-continued across $\{\Sigma=0\}$ by usual arguments.

Turning our attention, first, to the region where $r>0$, the occurrence of zeros of $\Delta_r$ requires that
$$\mu\ge \mu_c(a,e,\Lambda)>0
 \;.
$$
Hence, there is a \strictlyx  positive threshold for the mass of a black hole at given $a$ and $e$.
The solution with $\mu=\mu_c$ has the property that  $\Delta_r$ and its $r$-derivative have a joint zero, and can thus be found by equating to zero the resultant of these two polynomials in $r$. An explicit formula for $m_c=\Xi\mu_c$  can be given, which
takes a relatively simple form when expressed in terms of suitably renormalised parameters. We set
\beaa
 \alpha = \sqrt{\frac{|\Lambda|}{3}} a
 \quad
 &
 \Longleftrightarrow
 &
 \quad
a=\alpha \sqrt{\frac{3}{|\Lambda|}}
 \;,
\\
 \gamma = 9\frac{\alpha ^2+  \frac{|\Lambda|}{3} q^2 }{\left(1+\alpha ^2\right)^2}
 \quad
 &
 \Longleftrightarrow
 &
 \quad
 q^2 := \Xi  e^2 = \frac{3}{|\Lambda|}\left(\left(\frac{1+\alpha^2}{3}\right)^2\gamma-\alpha^2\right)
 \;,
\\
 \beta = \frac{3\sqrt{|\Lambda|}}{(1+\alpha^2)^{3/2}}\mu\Xi
 \quad
 &
 \Longleftrightarrow
 &
 \quad
  m := \Xi \mu
   = \frac{ (1+\alpha^2)^{3/2}}{3\sqrt{|\Lambda|}}\beta
 \;.
\eeaa
Letting $\beta_c$ be the value of $\beta$ corresponding to $\mu_c$, one finds
\bean
\lefteqn{
 \beta_c = \frac{\sqrt{-9+36 \gamma +\sqrt{3} \sqrt{(3+4 \gamma )^3}}}{3 \sqrt{2}}
 }
 &&
\\
 &&
 \Longleftrightarrow
 \quad
m_c^2=\frac{\left(1+\alpha ^2\right)^3 \left(-9+36 \gamma
+\sqrt{3} \sqrt{(3+4 \gamma )^3}\right)}{162 |\Lambda| }
 \;.
\eeal{28VII12.1}
When $q=0$, the graph of $\beta_c$ as a function of $\alpha$
can be found in Figure~\ref{F22VII12.1aswf}.
In general, the graph of $\beta_c$  as a function of $a$ and $q$ can be found in Figure~\ref{F30VII12.1aswfv}.

Note that if $q=0$, then $\gamma$ can be used as a replacement for $a$; otherwise, $\gamma$ is a substitute for $q$ at fixed $a$.

When $e=0$ we have $m_c=a+O(a^3)$ for small $a$, and $m_c\to  \frac{8}{3\sqrt{|\Lambda|}}$ as $|a|\nearrow \sqrt{|3/\Lambda|}$.

\begin{figure}
\begin{center}
\includegraphics[scale=1.3]{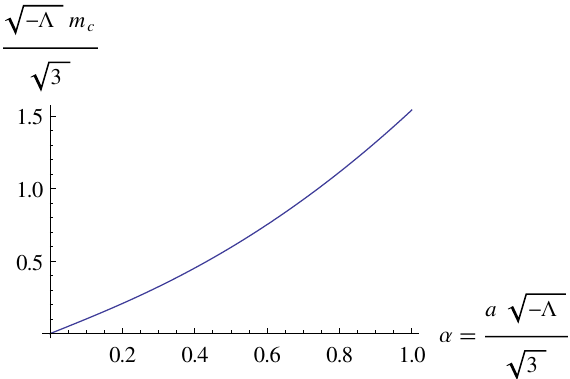}
\caption{The critical mass parameter $m_c\sqrt{| \Lambda/3|}=\Xi \mu_c \sqrt{|3/\Lambda|}$ as a function of  $|a| \sqrt{| \Lambda/3|}$ when $q=0$.}
\label{F22VII12.1aswf}
\end{center}
\end{figure}
\begin{figure}
\begin{center}
\includegraphics[scale=1]{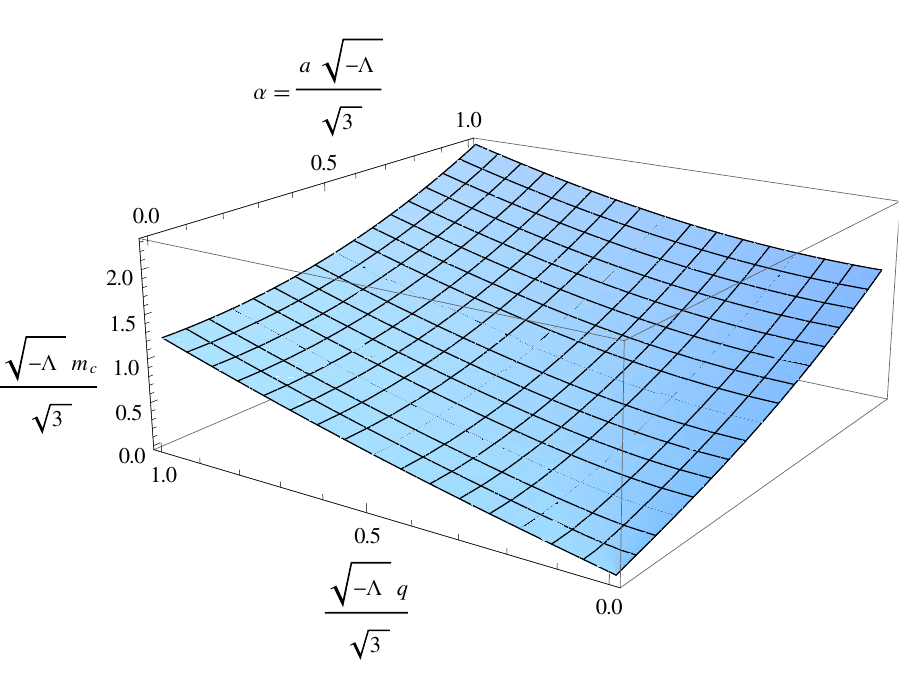}
\caption{The critical mass parameter $m_c\sqrt{\frac{|\Lambda|}{3}}$ as a function of  $\alpha=a \sqrt{\frac{|\Lambda|}{3}}$ and $q \sqrt{\frac{|\Lambda|}{3}}$.}
\label{F30VII12.1aswfv}
\end{center}
\end{figure}

According to \cite{HT}, the physically relevant mass of the solution is $\mu$ and \emph{not} $m$; because of the rescaling involved, we have $\mu_c\to\infty$ as $|a|\nearrow \sqrt{|3/\Lambda|}$.

We have $d^2 \Delta_r/dr^2 >0$, so that the set $\{\Delta_r \le 0\}$ is an interval $(r_-,r_+)$, with $0< r_-<r_+$.

  It follows from \eq{22VII12.5} that  $g_{\varphi\varphi}/\sin^2 (\theta)$ is \strictlyx  positive for $r>0$, and the analysis of the time-machine set is identical to the case $\Lambda>0$ as long as $\Xi>0$, which is assumed. We note that stable causality of each region  on which $\Delta_r$ has constant sign follows from \eq{22VII12.6} or \eq{22VII12.7}.

The projection metric is formally identical to that derived in Section~\ref{S2VII12.5}, with projection diagrams seen in Figure~\ref{F22VII12.1}.
\begin{figure}
\begin{center}
{\includegraphics[scale=0.9,angle=0]{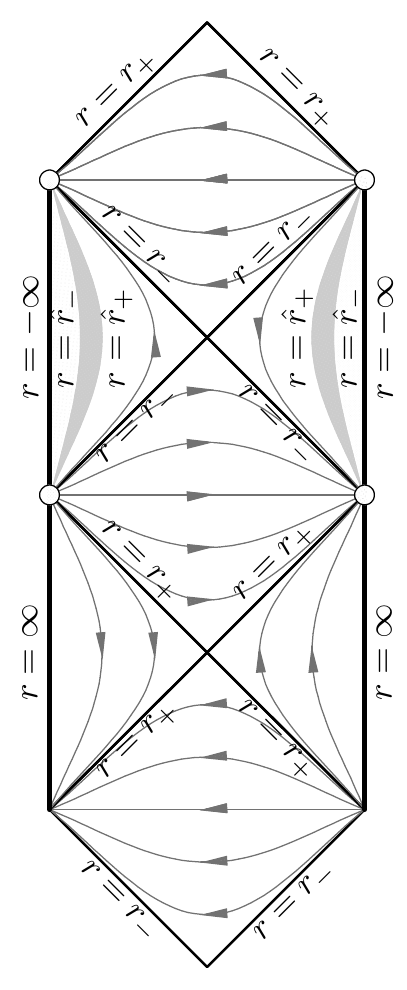}}
\hspace{1cm}
{\includegraphics[scale=0.9,angle=0]{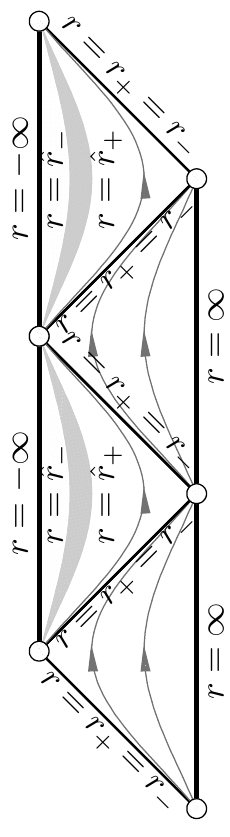}}
\caption{The projection diagrams for the Kerr-Newman - anti de Sitter metrics with two distinct zeros of $\Delta_r$ (left diagram) and one double zero (right diagram); see Remark~\ref{R25VII12.1}.}
\label{F22VII12.1}
\end{center}
\end{figure}

\subsection{The Emparan-Reall metrics}
 \label{S2VII12.9}

We consider the Emparan-Reall black-ring metric as presented in~\cite{EmparanReallLR}:
\beqa\label{metric}
ds^2&=&-\frac{F(y)}{F(x)}\left(dt-C\: R\:\frac{1+y}{F(y)}\:
d\psi\right)^2\nonumber\\[2mm]
&&+\frac{R^2}{(x-y)^2}\:F(x)\left[
-\frac{G(y)}{F(y)}d\psi^2-\frac{dy^2}{G(y)}
+\frac{dx^2}{G(x)}+\frac{G(x)}{F(x)}d\phi^2\right]\,,
\phantom{xxx}
\eeqa
where
\beq\label{fandg}
F(\xi)=1+\lambda\xi,\qquad G(\xi)=(1-\xi^2)(1+\nu\xi)\,,
\eeq
and
\beq\label{coeff}
C=\sqrt{\lambda(\lambda-
\nu)\frac{1+\lambda}{1-\lambda}}\,.
\eeq
The parameter $\lambda$ is chosen to be
\beq\label{noconical}
\lambda=\frac{2\nu}{1+\nu^2}\;,
\eeq
with the parameter $\nu$ lying in $(0,1)$, so that
\beq\label{lanurange}
0< \nu<\lambda<1\,.
\eeq
The coordinates $x$, $y$ lie in the ranges
$-\infty \leq y \leq -1$,  $-1 \leq x \leq 1$, assuming further that $(x,y)\ne (-1,-1)$.
The event horizons are located at $y=y_h=-1/\nu$ and the ergosurface is at $y=y_e=-1/\lambda$.
 The $\partial_\psi$--axis is at $y=-1$ and the $\partial_\phi$--axis is split into two parts $x=\pm 1$. Spatial infinity ${\it i}^0$ corresponds to $x=y=-1$.
 The metric becomes singular as $y\to-\infty$.

Although this is not immediately apparent from the current form of the metric, it is known~\cite{EmparanReall} that $\partial_\psi$ is spacelike or vanishing in the region of interest, with $g_{\psi\psi}>0$ away from the rotation axis $y=-1$.
Now, the metric (\ref{metric}) may be rewritten in the form
\begin{eqnarray}
g & = &
 \left(g_{tt}-\frac{g_{t\psi}^2}{g_{\psi\psi}}\right)dt^2-\frac{R^2}{(x-y)^2}\frac{F(x)}{G(y)}dy^2 \nonumber
 \\
 & &
  + \underbrace{g_{\psi\psi}\left(d\psi + \frac{g_{t\psi}}{g_{\psi\psi} } d t\right)^2 + g_{xx}dx^2 + g_{\phi\phi} d\phi^2}_{\ge 0}
 \;.
  \label{newmetric}
\end{eqnarray}
We have
\beq\label{e1}
g_{tt}-\frac{g_{t\psi}^2}{g_{\psi\psi}}=-\frac{G(y)F(y)F(x)}{F(x)^2G(y)+C^2(1+y)^2(x-y)^2}\;.
\eeq
It turns out that there is a non-obvious factorization of the denominator as
$$
 F(x)^2G(y)+C^2(1+y)^2(x-y)^2 = -F(y)I(x,y)
 \;,
$$
where $I$ is a second-order polynomial in $x$ and $y$ with coefficients depending upon $\nu$, sufficiently complicated so that it  cannot be usefully displayed here.  The polynomial $I$ turns out to be non-negative, which can be seen using a trick similar to one  in~\cite{CSPS}, as follows: One introduces new, non-negative, variables and parameters $(X,Y,\sigma)$ via the equations
\bel{28VI12.11}
 x=  X-1\;,\quad y= -Y-1\;,\quad \nu=\frac 1 {1+\sigma}\;,
\ee
with $0\leq X\leq 2$, $0\leq Y<+\infty$, $0<\sigma<+\infty$. A {\sc Mathematica} calculation shows that in this parameterization the function  $I $ is a rational function of the new variables, with a simple denominator which is explicitly non-negative, while the numerator is a complicated polynomial in $X$, $Y$, $\sigma$ with, however, \emph{all  coefficients positive}.

Let $\Omega=(x-y)/\sqrt{F(x)}$, then the function
\beq\label{IOm}
\kappa(x,y):=\Omega^2\left(g_{tt}-\frac{g_{t\psi}^2}{g_{\psi\psi}}\right)=-\frac{G(y)F(y)}{\frac{F(x)^2}{(x-y)^2}G(y)+C^2(1+y)^2}\;
\eeq
has extrema in $x$ only for $x=y=-1$ and $x=-1/\lambda<-1$. This may be seen from its derivative with respect to $x$,
which is explicitly  {non-positive} in the ranges of variables of interest:
\beq
 \frac{\partial \kappa}{\partial x} =
 -\frac{2G(y)^2F(y)^2F(x)(x-y)}{(F(x)^2G(y)+C^2(1+y)^2(x-y)^2)^2} = -\frac{2G(y)^2 F(x)(x-y)}{I(x,y)^2}
 \;.
\eeq
Therefore,
$$
  \frac{ (1+y)^2 G(y)}{I(-1,y)} = \kappa(-1,y) \ge \kappa(x,y) \ge \kappa (1,y)
   =\frac{ (1-y)^2 G(y)}{I(1,y)}
 \;.
$$
Since both $I(-1,y)$ and $I(1,y)$ are positive,   in the domain of outer communications $\{-1/\nu<y\le -1\}$ where  $G(y)$ is negative  we obtain
\beq
 \frac{ -G(y)(1+y)^2}{I(-1,y) }
  \leq\left|\Omega^2\left(g_{tt}-\frac{g_{t\psi}^2}{g_{\psi\psi}}\right)\right|\leq
   \frac{ -G(y)(1-y)^2}{I(1,y)}\;.
\eeq
One finds
$$
 I(1,y) =\frac{1 + \lambda}{1-\lambda} (-1 + y^2) (1 - y (\lambda - \nu) - \lambda \nu)
  \;,
$$
which leads to the projection metric
\beq
\gamma:= \chi(y) \frac{G(y)}{(-1-y)}dt^2-\frac{R^2}{G(y)}dy^2\;,
\eeq
where, using the variables \eq{28VI12.11} to make manifest the positivity of $\chi$ in the range of variables of interest,
\beaa
  \chi(y)
 &= &
  \frac{(1-y)(1-\lambda) }{(1+\lambda)(1 - y (\lambda - \nu) - \lambda \nu)}
\\
 &= &
  \frac{(2 + Y) \sigma (1 + \sigma) (2 + 2 \sigma + \sigma^2) }{  (2 + \sigma)^3 (2 + Y + \sigma)} >0
 \;.
\eeaa
The calculation of \eq{Krakgen2VII12.a2} leads to the following conformal metric
\bea \label{9VII12.1}
 &
  \displaystyle \twogL=R\sqrt{\frac\chi{|1+y|}} \left( -\hat F  dt^{2}+\hat F^{-1} dr^{2}\right) \;,
 \ \mbox{where $\hat F = -\frac 1 R \sqrt{\frac\chi {{|1+y|}} }G$}
 \;.
 &
\eea
Since the integral of $\hat F^{-1}$ diverges at the event horizon, and is finite at $y=-1$ (which corresponds \emph{both} to an axis of rotation and the asymptotic region at infinity), the
analysis in Section~\ref{S2VII12.1} shows that the corresponding projection diagram is as in Figure~\ref{F25VII12.99}.
\begin{figure}
\begin{center}
{\includegraphics[scale=1]{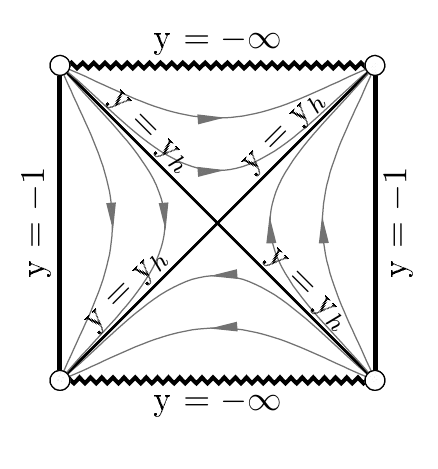}}
\caption{The projection diagram for the Emparan-Reall black rings. The arrows indicate the causal character of the orbits of the isometry group. The boundary $y=-1$ is covered, via the projection map, by the axis of rotation \emph{and} by spatial infinity $i^0$. Curves approaching the conformal null infinities $\scripm$ asymptote to the missing corners in the diagram. {\color{red} }
\label{F25VII12.99}}
\end{center}
\end{figure}

It is instructive to compare this to the projection diagram for five-dimensional Minkowski space-time
$$(t,\hat r \cos \phi, \hat r \sin \phi, \tilde r \cos \psi, \tilde r \sin \psi)\equiv (t,\hat x,\hat y, \tilde x,\tilde y)\in \R^5
$$
parameterized by ring-type coordinates:
$$
y= -\frac{\hat r^2}{\left(\hat r^2+\tilde r^2\right)^2}-1
\;,
\quad
 x =  \frac{\tilde r^2}{\left(\hat r^2+\tilde r^2\right)^2}-1
 \;,
 \quad
 \hat r = \sqrt{\hat x^2 + \hat y ^2}
 \;,\quad \tilde r = \sqrt{\tilde x^2 + \tilde y ^2}
 \;.
$$
For fixed $x\ne 0$, $ y\ne 0$ we obtain a torus as $\varphi $ and $\psi$ vary over $S^1$.
The image of the resulting map is the set $x\ge-1$, $y\le -1$, $(x,y) \ne (-1,-1)$. Since
$$
 x-y = \frac 1 {\hr^2+\tilde r^2}
 \;,
$$
the spheres $\hr^2 +\tilde r^2=:r^2=\const$ are mapped to subsets of the lines $x=y + 1/r^2$, and the limit $r\to \infty$ corresponds to $0\le x-y \to 0$ (hence $x\to -1$ and $y\to -1$).
The inverse transformation reads
$$
\hat r = \frac{\sqrt{-y-1}}{x-y} \;,
 \quad
 \tilde r = \frac{\sqrt{x+1}}{x-y}
 \;.
$$
The Minkowski metric takes the form
\beaa
 \eta
  &  =  &
   -dt^2 + d\hat x ^2 + d\hat  y ^2 + d\tilde x ^2 + d\tilde y^2
\\
  & = &
    -dt^2 + d\hat r ^2 +  \hat  r ^2 d\varphi^2 + d\tilde r ^2 +  \tilde r^2 d \psi^2
\\
  & = &
    -dt^2 +  \frac{   {dy}^2}{4   (-y-1) (x-y)^2} + \underbrace{\frac{ {dx}^2 }{4 (x+1) (x-y)^2} +  \hat  r ^2 d\varphi^2   +  \tilde r^2 d \psi^2}_{\ge 0}
    \;.
\eeaa
 Thus, for any $\eta$-causal vector $X$,
\beaa
 \eta(X,X) & \ge & -(X^t)^2  + \frac{   {(X^y)}^2}{4   (-y-1) (x-y)^2}
  \;.
\eeaa
There is a problem with the right-hand side since, at fixed $y$, $x$ is allowed to go to infinity,  and so there is no \strictlyx  positive lower bound on the coefficient of $(X^y)^2$. However, if we restrict attention to the set %
$$
 r=\sqrt{\hat r^2 + \tilde r^2} \ge R
$$
for some $R>0$, we obtain
\beaa
 \eta(X,X) & \ge & -(X^t)^2  + \frac{  R^4 {(X^y)}^2}{4   (-y-1) }
  \;.
\eeaa
This leads to the conformal projection metric, for {$-1 -\frac 1 {R^2} =:y_R \le y\le -1$},
\bean
 \gamma &:= & -dt^2 +\frac{  R^4 dy^2}{4  |y+1| }
\\
 & = & -dt^2 + \left( d \left(R^2  \sqrt{|y+1|}\right)\right)^2
 \nonumber
\\
  &= &   \frac {R^2} {2\sqrt{|y+1| }} \left( -  \frac  {2\sqrt{|y+1| }} {R^2} dt^2 + \frac {R^2} {2\sqrt{|y+1| }} dy^2\right)
 \;.
\eeal{6VII12.1}
Introducing a new coordinate $y'= -R^2\sqrt{-y-1}$ we have
\beaa
\gamma=-dt^2+dy'^2\;,
\eeaa
where $-1\le y' \le 0$. Therefore, the projection diagram corresponds to a subset of the standard diagram for a two-dimensional Minkowski spacetime, see Figure~\ref{F25VII12.98}.
\begin{figure}
\begin{center}
{\includegraphics[scale=1,angle=90,trim = 2.5mm 0mm 0mm 0mm, clip]{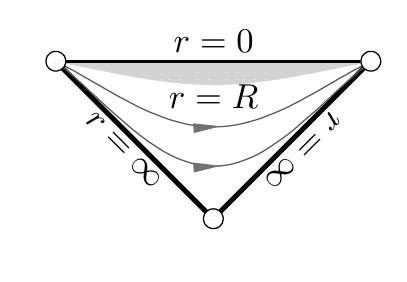}}
{\includegraphics[scale=1,angle=90]{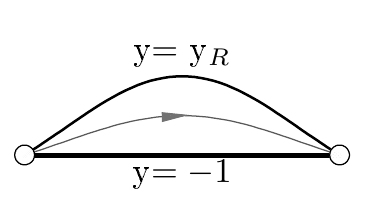}}
\caption{The projection diagram for the complement of a world-tube $\R\times B(R)$ in five-dimensional Minkowski space-time using spherical coordinates (left figure, where the shaded region has to be removed), or using ring coordinates (right figure). In the right figure the right boundary $y=-1$ is covered, via the projection map,  both by the axis of rotation and by spatial infinity, while null infinity projects to the missing points at the top and at the bottom of the diagram.
 {\color{red}  }
\label{F25VII12.98}}
\end{center}
\end{figure}

\subsection{The Pomeransky-Senkov metrics}
 \label{S2VII12.10}

We consider the \emph{Pomeransky-Senkov metrics}~\cite{PS},
\begin{eqnarray}
g &= & \frac{2 H(x,y)k^2}
{(1-\nu)^2(x-y)^2}\left(\frac{dx^2}{G(x)}-
\frac{dy^2}{G(y)}\right)  -2\frac{ J(x,y)}{H(y,x)}d\varphi d\psi\nonumber\\
&&-
\frac{H(y,x)}{H(x,y)}(dt+\Omega)^2-\frac{F(x,y)}{H(y,x)}d\psi^2+
\frac{F(y,x)}{H(y,x)}d\varphi^2\;,
\label{eq:line-element1}
\end{eqnarray}
where $\Omega$ is a 1-form given by
$$
\Omega=M(x,y)d\psi+P(x,y)d\varphi
 \;.
$$
The definitions of the metric functions may be found in \cite{PS}.\footnote{We use $(\psi,\varphi)$ where Pomeransky \& Senkov use $(\varphi,\psi)$.}
The metric depends on three constants: $k$, $\nu$, $\lambda$, where $k$ is assumed to be in $\R^*$, while the
parameters $\lambda$ and $\nu$ are restricted to the set\footnote{$\nu=0$ corresponds to Emparan-Reall metric which has been already analyzed in Section~\ref{S2VII12.9}.}
\bel{3IX.1}
 \{(\nu,\lambda):\ \nu\in (0,1)\;,\
2\sqrt{\nu}\le \lambda< 1 +\nu\}
  \;.
\ee
The coordinates $x$, $y$, $\varphi$,
$\psi$, and  $t$ vary within the ranges $-1\leq x\leq 1$,
$-\infty<y<-1$, $0\leq\varphi\leq 2\pi$, $0\leq\psi\leq 2\pi$ and
$-\infty<t<\infty$.

A  Cauchy horizon is located at
$$
y_c:=-\frac{\lambda+\sqrt{\lambda^2-4\nu}}{2\nu}\;,
$$
and the event horizon corresponds to
$$
y_h:=-\frac{\lambda-\sqrt{\lambda^2-4\nu}}{2\nu}\;.
$$
Using an appropriate Gauss {diagonalization}, the metric may be rewritten in the form
\begin{eqnarray}\nonumber
&&
g=\overbrace{\frac{g_{t\psi}^2 g_{\varphi\varphi} - 2 g_{t\varphi} g_{t\psi} g_{\psi\varphi} +
   g_{t\varphi}^2 g_{\psi\psi} +
   g_{tt} (g_{\psi\varphi}^2 -
      g_{\varphi\varphi} g_{\psi\psi})}{g_{\psi\varphi}^2 -
 g_{\varphi\varphi} g_{\psi\psi}}}^{(*)}dt^2+g_{yy}dy^2 \\
 &&
+\underbrace{g_{xx}dx^2+
\left(g_{\varphi\varphi} -
   \frac{g_{\psi\varphi}^2}{g_{\psi\psi}}\right)\left(
d\varphi + \frac{g_{t\varphi} - \frac{g_{t\psi} g_{\psi\varphi}}{g_{\psi\psi}}}{
   g_{\varphi\varphi} - \frac{g_{\psi\varphi}^2}{g_{\psi\psi}}}dt
\right)^2 +\frac{(g_{t\psi}dt +  g_{\psi\varphi}d\varphi +
  g_{\psi\psi}d\psi)^2}{g_{\psi\psi}}}_{(**)}
  \;.
  \nonumber\\\label{PSmetric}
\end{eqnarray}
 The positive-definiteness of $(**)$ for $y>y_c$ follows from \cite{CSPS,CCGP}.
Note that $g_{\psi\psi}<0$ would give a timelike Killing vector $\partial_\psi$, and that
$g_{\varphi\varphi}{g_{\psi\psi}} -
    {g_{\psi\varphi}^2} <0$ would lead to some combination of the periodic Killing vectors $\partial_\varphi$ and $\partial_\psi$ being timelike, so the term
 $(**)$ in \eq{PSmetric} is non-negative on any region where there are no obvious causality violations.

The coefficient $(*) $ in front of $dt^2$ is negative for $y>y_h$ and positive for $y<y_h$, vanishing at $y=y_h$. This may be seen in the reparameterized form of the Pomeransky-Senkov solution that was introduced in \cite{CSPS}: Indeed, let $a$, $b$ be the new coordinates as in \cite{CSPS} replacing $x$ and $y$, respectively, and let us reparameterize $\nu$, $\lambda$  by $c$, $d$ again as in \cite{CSPS}, where all the variables $a$, $b$, $c$, $d$ are non-negative above the Cauchy horizon, $y>y_c$:
\begin{eqnarray}\nonumber
 x &=& -1 + \frac 2 {1+a}\;,\\\nonumber
 y &=&  -1 - \frac{d (4 + c+2 d )}{(1 + b) (2 + c)}\;,\\\nonumber
\nu &=& \frac 1{(1 + d)^2}\;,\\
\lambda  &=& 2 \frac{2 d^2 + 2 (2 + c) d+ (2 + c)^2}{(2 + c) (1 + d) (2 + c + 2 d )}\;.
 \label{14VIII12.11}
\end{eqnarray}
Set
\begin{eqnarray}
\kappa&:=&(*)\,\Omega^2\;,\\
\Omega^2&:=&\frac{(x - y)^2 (1 - \nu)^2}{2k^2H(x, y)}\;.
\end{eqnarray}
Using {\sc Mathematica} one finds that $\kappa$ takes the form
$$
\kappa=-\Omega^2(y-y_h) Q\;,
$$
where $Q=Q(a,b,c,d)$ is a huge rational function in $(a,b,c,d)$ with \emph{all coefficients positive}. To obtain the corresponding projection metric $\gamma$ one would have, e.g., to find sharp lower and upper bounds for $Q$, at fixed $y$, which would lead to
$$
 \gamma := - (y-y_h) \sup_{\mbox{\rm \scriptsize y fixed}} |Q|\ dt^2 - \frac 1 { G(y)} dy^2
 \;.
$$
This requires analyzing a complicated rational function, which we have not been able to do so far. We hope to return to this issue in the future.

We expect the corresponding projection diagram to look like that for Kerr - anti de Sitter space-time of Figure~\ref{F22VII12.1}, with $r=\infty$ there replaced by $y=-1$, $r=-\infty$ replaced by $y=1$ with an appropriate analytic continuation of the metric to positive $y$'s (compare~\cite{CCGP}), $r_+$ replaced by $y_h$ and $r_-$ replaced by $y_c$. The shaded regions in the negative region there might be non-connected for some values of parameters, and always extend to the boundary at infinity in the relevant diamond~\cite{CCGP}.

Recall that a substantial part of the work in~\cite{CCGP} was to show that the function $H(x,y)$ had no zeros for $y>y_c$. We note that the reparameterization
$$
y \rightarrow -1 - \frac{cd}{(1 + b) (2 + c + 2 d)}
$$
 of \cite{CSPS}
  (with the remaining formulae \eq{14VIII12.11} remaining the same), gives
 $$
 H(x,y)= \frac{P(a,b,c,d)}{(1+a)^2 (1+b)^2 (2+c)^2 (1+d)^6 (2+c+2d)^4}
 \;,
 $$
where $P$ is a huge polynomial with \emph{all coefficients positive} for $y>y_h$. This establishes immediately positivity of $H(x,y)$ in the domain of outer communications. We have, however, not been able to find a simple proof of positivity of $H(x,y)$ in the whole range $y>y_c$.

\section{An application to spatially compact $\Uone\times\Uone$ symmetric models with compact Cauchy horizons}
 \label{S14VIII12.1}

 In this section we wish to  use the Kerr-Newman - (a)dS family of metrics to construct explicit examples of maximal, four-dimensional, $\Uone\times\Uone$  symmetric, electrovacuum or vacuum models, with or without cosmological constant, containing a spatially compact partial Cauchy surface. Similarly, five-dimensional,  $\Uone\times\Uone\times \Uone$  symmetric, spatially compact vacuum models with spatially compact partial Cauchy surfaces can be constructed using the Emparan-Reall or Pomeransky-Senkov metrics. We will show how the projection diagrams constructed so far can be used to understand maximal (non-globally hyperbolic) extensions of the maximal globally hyperbolic regions in such models, and for the Taub-NUT metrics.

\subsection{Kerr-Newman-(a)dS-type and Pomeransky-Senkov-type models}
 \label{ss14VIII12.1}

The diamonds and triangles which have been used to construct our diagrams so far  will be referred to as \emph{blocs}. Here the notion of a triangle is understood \emph{up to diffeomorphism}, thus planar sets with three corners, connected by smooth curves intersecting only at the corners which are \emph{not} necessarily straight lines, are also considered to be triangles.

In the interior of each bloc  one can  periodically identify points lying along the orbits of the action of the $\R$ factor of the isometry group. Here we are only interested in the connected component of the identity of the group, which is $\R\times \Uone$ in the four-dimensional case, and $\R\times \Uone\times\Uone$ in the five-dimensional case.

Note that isometries of space-time extend smoothly across all bloc boundaries. For example, in the coordinates $(v,r,\theta,\tilde\varphi)$ discussed in the paragraph around \eq{16VIII12.1x}, p.~\pageref{16VIII12.1x}, translations in $t$ become translations in $v$; similarly for the $(u,r,\theta,\tilde\varphi)$ coordinates.
Using the $(U,V,\theta,\tilde \varphi)$ local coordinates near the intersection of two Killing horizons, translations in $t$ become boosts in the $(U,V)$ plane.

Consider  one of the blocs, out of any of the diagrams constructed above,  in which the orbits of the isometry group  are spacelike. (Note that no such diamond or triangle has a shaded area which needs to be excised, as the shadings occur only within those building blocs where the  isometry orbits are timelike.) It can be seen that the periodic identifications result then in a spatially compact maximal globally hyperbolic space-time with $S^1\times S^2$ spatial topology, respectively with $ S^1\times S^1 \times S^2$ topology.

Now, each  diamond in our diagrams has four null boundaries which naturally split into pairs, as follows:
In each bloc in which the isometry orbits are spacelike, we will say that two boundaries are \emph{orbit-adjacent} if  both boundaries lie to the future of the bloc, or both to the past. In a bloc where the isometry orbits are timelike, boundaries will be said \emph{orbit-adjacent} if they are both to the left or both to the right.

One out of each pair of orbit-adjacent null boundaries of a bloc with spacelike isometry-orbits corresponds, in the periodically identified space-time, to a compact Cauchy horizon across which the space-time can be continued to a periodically identified adjacent bloc. Which of the two adjacent boundaries will become a Cauchy horizon is a matter of choice; once such a choice has been made, the other boundary \emph{cannot} be attached anymore: those  geodesics which, in the unidentified space-time, would have been crossing the second boundary become, in the periodically identified space-time, incomplete inextendible geodesics. This behaviour is well known from  Taub-NUT space-times~\cite{Taub,ChImaxTaubNUT,Misner}, and is easily seen as follows:

Consider a sequence of points $p_i:=(  t_i, r_i)$ such that $p_i$ converges to a point $p$ on a horizon  in a projection diagram in which no periodic identifications have been made. Let $T>0$ be the period with which the points are identified along the isometry orbits, thus for every  $n\in \Z$ points $(t,r)$ and $(t+nT,r)$ represent  the same point of the quotient manifold. It should be clear from the form of the Eddington-Finkelstein type coordinates $u$ and $v$ used to perform  the two distinct extensions (see the paragraph around \eq{16VIII12.1x}, p.~\pageref{16VIII12.1x}) that there exists a sequence $n_i\in \Z$ such that, passing to a subsequence if necessary, the sequence $q_i=(t_i+n_iT, r_i)$ converges to some point $q$ in the companion orbit-adjacent boundary, see Figure~\ref{F18VIII12.1}.
\begin{figure}
\begin{center}
{\includegraphics[scale=1.8,angle=0, trim= 0mm 4mm 0mm 3mm,clip]{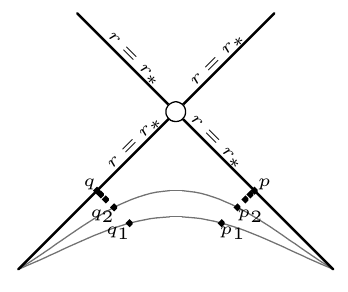}}%
\caption{The sequences $q_i$ and $p_i$. Rotating the figure by integer multiples of 90 degrees shows that the problem of non-unique limits arises on any pair of orbit-adjacent boundaries. {\color{red}}}
\label{F18VIII12.1}
\end{center}
\end{figure}

Denote by $[p ]$ the class of $p$ under the equivalence relation $(t,r)\sim (t+nT,r)$, where $n\in \Z$ and $T$ is the period.
Suppose  that one could construct simultaneously an extension of the quotient manifold across both orbit-adjacent boundaries. Then the sequence of points $[q_i]=[p_i]$  would have two distinct points $[p]$ and $[q]$ as limit points, which is not possible. This establishes our claim.

Returning to our main line of thought, note that a  periodically identified  building bloc in which the isometry orbits are timelike  will have obvious causality violations throughout, as a linear combination of the periodic Killing vectors becomes timelike there.

The branching construction, where one out of the pair of orbit-adjacent boundaries is chosen to perform the extension, can be continued at each  bloc in which the isometry orbits are spacelike.
This shows that maximal extensions are obtained from any connected union of blocs such that in each bloc an extension is carried out across precisely one out of each pair of orbit-adjacent boundaries.
  Some such subsets of the plane might only comprise a finite number of blocs, as seen trivially in Figure~\ref{F25VI12.7}. Clearly an infinite number of distinct finite, semi-infinite, or infinite sequences of blocs can be constructed in the diagram of Figure~\ref{F25VI12.4}. Two sequences of blocs which are not related by one of the discrete isometries of the diagram will lead to non-isometric maximal extensions of the maximal globally hyperbolic initial region.

\subsection{Taub-NUT metrics}
 \label{ss14VIII12.2}

We have seen at the end of Section~\ref{S14VIII12.3} how to construct a projection diagram for Gowdy cosmological models. Those models all contain $\Uone\times\Uone$ as part of their isometry group. The corresponding projection diagrams constructed in Section~\ref{S14VIII12.3} were obtained by projecting out the isometry orbits. This is rather different from the remaining projection diagrams constructed in this work, where only one of the  coordinates along the Killing orbits was projected out.

It is instructive to carry out explicitly both procedures for the Taub-NUT metrics, which belong to the Gowdy class.
Using
Euler angles $ (\zeta ,\theta,\varphi)$ to parameterize $S^3$,
the Taub-NUT  metrics~\cite{Taub,NUT} take  the form
 \be
 g= -U^{-1}dt^2 +(2\ell )^2U(d\zeta  + \cos(\theta)\, d\varphi)^2 +
(t^2 + \ell ^2)(d\theta^2+\sin^2(\theta)\, d\varphi^2)\,.
\label{14VIII12.1}
\ee
Here
$$U(t)=  -1 +
{2(mt + \ell ^2)\over
  t^2 + \ell ^2}=
\frac {(t_+-t)(t-t_-)}{t^2+\ell ^2}\;,$$
with
$$ t_\pm := m\pm
\sqrt{m^2+\ell ^2}\;.$$
Further, $\ell $ and $m$ are real numbers with $\ell  > 0$. The region   $\{t\in (t_-,t_+)\}$ will be referred to as the \emph{Taub space-time}.

The metric induced on the sections $\theta=\const$, $\varphi = \const'$,  of the Taub space-time reads
\bel{15VIII12.1}
 \gamma_0:=  -U^{-1}dt^2 +(2\ell )^2U d\zeta  ^2
 \;.
\ee
As already discussed by Hawking and Ellis~\cite{HE}, this is a metric to which the methods of Section~\ref{S2VII12.1} apply \emph{provided that the $4\pi$-periodic identifications in $\zeta$ are relaxed}.
\begin{figure}
\begin{center}
{\includegraphics[scale=.7,angle=0]{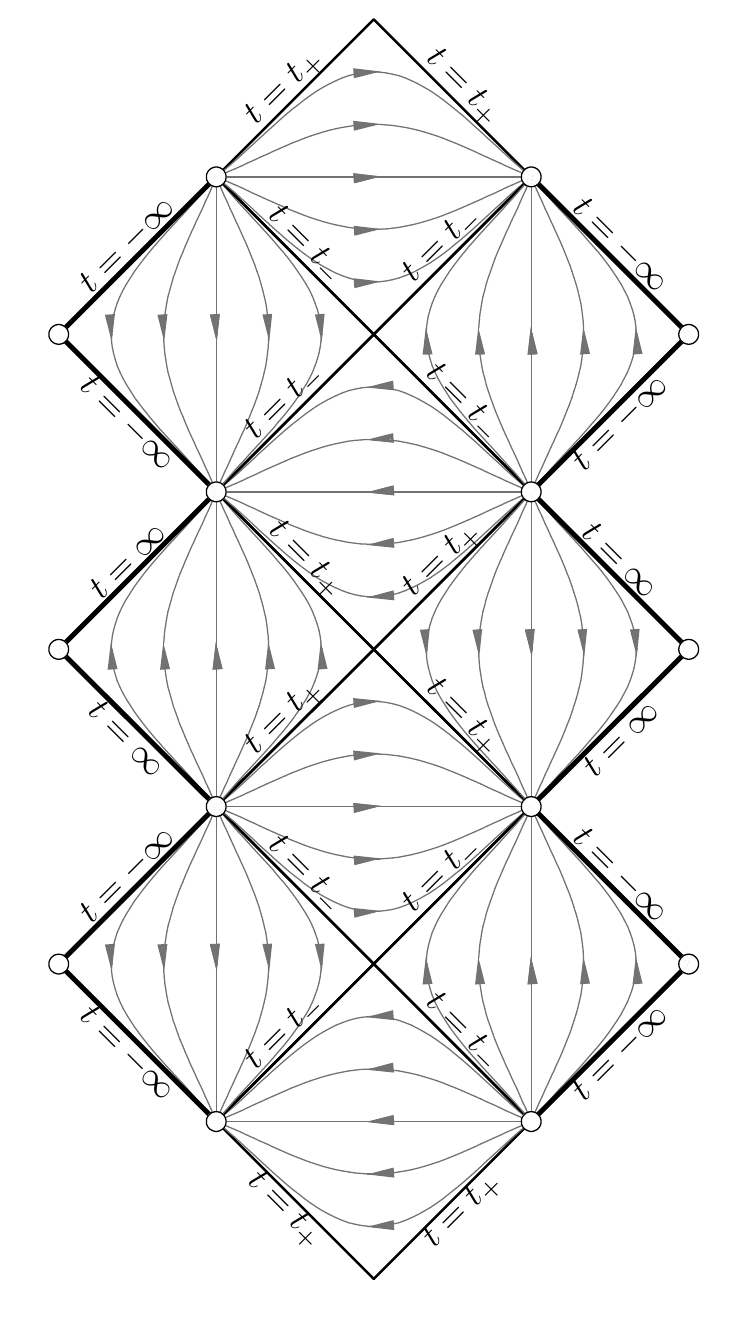}}%
{\includegraphics[scale=.7,angle=0,trim = 0mm -20mm 0mm 0mm]{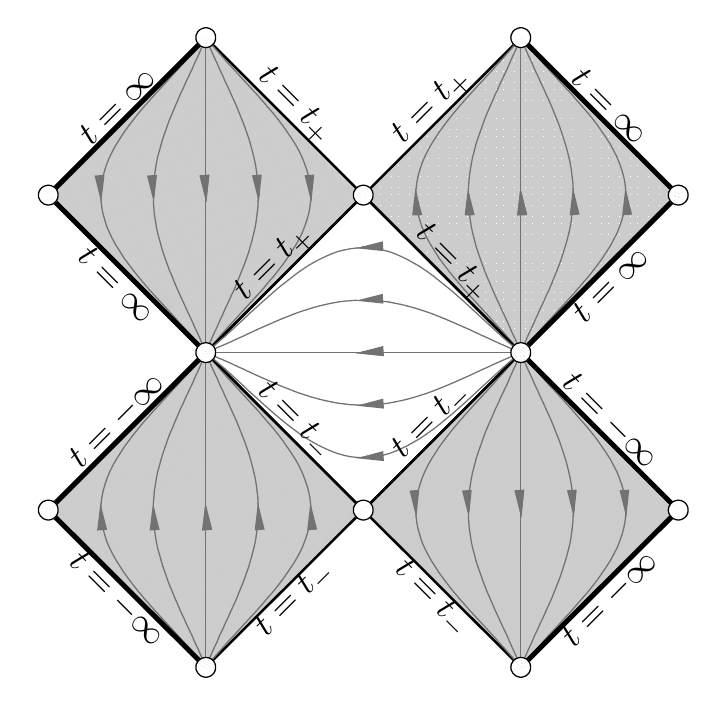}}%
\caption{The left diagram is the conformal diagram for an extension of the universal covering space of the sections $\theta=\const$, $\varphi = \const'$,  of the Taub space-time. The right diagram represents simultaneously the four possible diagrams for the maximal extensions, within the Taub-NUT class, with compact Cauchy horizons, of the Taub space-time.  After invoking the left-right symmetry of the diagram, which lifts to an isometry of the extended space-time, the four diagrams  lead to two non-isometric spacetimes. {\color{red}}}
\label{F16VIII12.1}
\end{center}
\end{figure}
Since $U$ has two simple zeros, and no singularities, the conformal diagram for the corresponding maximally extended two-dimensional space-time equipped with the metric $\gamma_0$  coincides with
the left diagram in Figure~\ref{F16VIII12.1},
compare~\cite[Figure~33]{HE}.
The discussion of the last paragraph of the previous section applies and, together with the left diagram in Figure~\ref{F16VIII12.1}, provides a family of simply connected maximal extensions of the sections $\theta=\const$, $\varphi = \const'$,  of the Taub space-time.

However, it is not clear how to relate the above to extensions of the four-dimensional space-time.
Note that projecting out the $\zeta$ and $\varphi$ variables in the region where $U>0$,  using the projection map $\pi_1(t,\zeta,\theta,\varphi):= (t,\theta)$, one is left with the two-dimensional metric
\be
 \gamma_1:=  -U^{-1}dt^2   + (t^2 + \ell ^2)\, d\theta^2
 \;,
\label{14VIII12.2}
\ee
which leads to the flat metric on the Gowdy square as the projection metric. (The coordinate $t$ here is  not the same as the Gowdy $t$ coordinate, but the projection diagram remains a square.)  And one is left wondering how this fits with the previous picture.

Now, one can attempt instead to project out the $\theta$ and $\varphi$ variables, with the projection map
\bel{15VIII12.12}
 \pi_2(t,\zeta,\theta,\varphi):= (t,\zeta)
 \;.
\ee
For this we note the trivial identity
\bel{15VIII12.5}
 g_{\zeta\zeta} d\zeta^2 +
 2g_{\varphi\zeta} d\varphi\, d\zeta+
 g_{\varphi\varphi} d\varphi^2
 =
 \big(g_{\zeta\zeta} - \frac{g_{\varphi \zeta}^2}{g_{\varphi\varphi}}\big) d\zeta^2 +
 \underbrace{g_{\varphi\varphi} \big(d\varphi +
 \frac{g_{\varphi \zeta}}{g_{\varphi\varphi} } d\zeta)^2 }_{(*)}
 \;.
\ee
Since the left-hand side is positive-definite on Taub space, where $U>0$, both $g_{\zeta\zeta} - \frac{g_{\varphi \zeta}^2}{g_{\varphi\varphi}}$ and $
 g_{\varphi\varphi} $ are non-negative there.
Indeed,
\beal{15VIII12.6}
g_{\varphi\varphi} & = &
\left(\ell^2+t^2\right) \sin ^2(\theta )+4 \ell^2 U \cos ^2(\theta
   )
   \;,
\\
 g_{\zeta\zeta} - \frac{g_{\varphi \zeta}^2}{g_{\varphi\varphi}}
  & = & (2\ell)^2\bigg(1 - \frac {(2\ell)^2 U \cos^2(\theta)}{g_{\varphi\varphi}}
  \bigg) U
   \nonumber
\\
 & =  & \underbrace{ \frac{4 \ell^2\left(\ell^2+t^2\right) \sin ^2(\theta
   )  }{\left(\ell^2+t^2\right) \sin ^2(\theta )+4 \ell^2 U \cos ^2(\theta
   )}}_{(**)} U
   \;.
\eeal{15VIII12.7}
However, perhaps not unsurprisingly given the character of the coordinates involved, the function $(**)$ in \eq{15VIII12.7}  does not have a positive lower bound independent of $\theta\in [0,2\pi]$, which is unfortunate for our purposes. To sidestep this drawback we   choose  a  number $0<\epsilon<1$ and restrict ourselves to the range $\theta\in [\theta_\epsilon,\pi-\theta_\epsilon]$, where $\theta_\epsilon\in[0,\pi/2]$ is defined by
$$
 \sin^2(\theta_\epsilon) =
 \epsilon
 \;.
$$
Now, $g_{\varphi\varphi}$ is  \strictlyx positive for large $t$, independently of $\theta$. Next, $g_{\varphi\varphi}$  equals $4\ell^2 U$ at the axes of rotation $\sin(\theta)=0$, and equals $\ell^2+t^2$ at $\theta=\pi/2$. Hence, keeping in mind that $U$ is monotonic away from $(t_-,t_+)$, for $\epsilon$
small enough there will exist values
$$ \hat t_\pm(\epsilon)\;, \ \mbox{with $\hat t_-(\epsilon)< t_- < 0 < t_+ <\hat t_+(\epsilon)$}
$$
such that $g_{\varphi\varphi}$ will be negative somewhere in the region $\big(\hat t_-(\epsilon),t_-\big)\cup \big(t_+,\hat t_+(\epsilon)\big)$, and will be positive outside of this region. We choose those numbers to be optimal with respect to those properties.

On the other hand, for $\epsilon$ close enough to $1$ the metric coefficient $g_{\varphi\varphi}$ will be  \strictlyx positive for all  $\theta\in [\theta_\epsilon,\pi-\theta_\epsilon]$ and $t<t_-$. In this case we set $\hat t_-(\epsilon)=t_-$, so that the interval $\big(\hat t_-(\epsilon),t_-\big)$ is empty. Similarly, there will exist a range  of $\epsilon$ for which $\hat t_+(\epsilon)=t_+$, and $ \big(t_+,\hat t_+(\epsilon)\big)=\emptyset$. The relevant ranges of $\epsilon$ will coincide only if $m= 0$.

We note
$$
 \partial_\theta\bigg( g_{\zeta\zeta} - \frac{g_{\varphi \zeta}^2}{g_{\varphi\varphi}}\bigg) = \frac{16 \ell^4 U^2 \left(\ell^2+t^2\right) \sin (2 \theta
   )}{\left(\left(\ell^2+t^2\right) \sin ^2(\theta )+4 \ell^2 U \cos
   ^2(\theta )\right)^2}
   \;,
$$
which shows that, for
\bel{15VIII12.10}
\mbox{$t\not\in \big(\hat t_-(\epsilon),t_-\big)\cup \big(t_+,\hat t_+(\epsilon)\big)$ and   $\theta\in (\theta_\epsilon,\pi-\theta_\epsilon)$,}
\ee
the multiplicative coefficient  $(**)$ of $U$ in \eq{15VIII12.7}
will satisfy
\bel{15VIII12.9}
(**) \ge
{ \frac{4 \ell^2\left(\ell^2+t^2\right) \sin ^2(\theta_\epsilon
   )  }{\left(\ell^2+t^2\right) \sin ^2(\theta_\epsilon )+4 \ell^2 U \cos ^2(\theta_\epsilon
   )}}=: f_\epsilon(t)
   \;.
\ee
We are ready now to construct the projection metric  in the region  \eq{15VIII12.10}.
Removing from the metric tensor \eq{14VIII12.1} the terms  $(*)$ appearing in \eq{15VIII12.5}, as well as the $d\theta^2$ terms, and using \eq{15VIII12.9} one finds, for $g$-causal vectors $X$,
$$
 g(X,X)\ge \gamma_2((\pi_2)_*X,(\pi_2)_*X)
 \;,
$$
with $\pi_2$ as in \eq{15VIII12.12}, and
where
\be
 \gamma_2:=  -U^{-1}dt^2 +f_\epsilon\, U d\zeta  ^2
 \;.
\label{14VIII12.3}
\ee
Since $U$ has exactly two simple zeros and is finite everywhere,
and for $\epsilon$ such that $g_{\varphi\varphi}$ is positive on the region $\theta\in [\theta_\epsilon,\pi-\theta_\epsilon]$, the projection diagram
for that region, \emph{in a space-time in which no periodic identifications in $\zeta$ are made}, is given by the left diagram of Figure~\ref{F16VIII12.1}. The reader should have no difficulties finding the corresponding diagrams for the remaining values of $\epsilon$.

However, we are in fact interested in those space-times where $\zeta$ is $4\pi$ periodic. This has two consequences: a) there are closed timelike Killing orbits in all the regions where $U$ is negative, and b) no simultaneous extensions are possible across two orbit-adjacent boundaries.  It then follows (see the right diagram of Figure~\ref{F16VIII12.1})
that there are,  within the Taub-NUT class, only two non-isometric, maximal, vacuum extensions across compact Cauchy horizons  of the Taub space-time.
(Compare~\cite[Proposition~4.5 and Theorem~1.2]{chrusciel:rendall:SCC} for the
local uniqueness of extensions, and \cite{ChConformalBoundary} for a discussion of extensions with non-compact Killing horizons.)

\bigskip

\noindent{\sc Acknowledgements.}
PTC was supported in part by Narodowe Centrum Nauki under the grant DEC-2011/03/B/ST/02625.
SJS was supported by the John Templeton Foundation and would like to thank the University of Vienna for hospitality.

\bibliographystyle{amsplain}%
\bibliography{../references/hip_bib,%
../references/reffile,%
../references/newbiblio,%
../references/newbiblio2,%
../references/chrusciel,%
../references/bibl,%
../references/howard,%
../references/bartnik,%
../references/myGR,%
../references/newbib,%
../references/Energy,%
../references/dp-BAMS,%
../references/prop2,%
../references/besse2,%
../references/netbiblio,%
../references/PDE}

\def\polhk#1{\setbox0=\hbox{#1}{\ooalign{\hidewidth
  \lower1.5ex\hbox{`}\hidewidth\crcr\unhbox0}}} \def\cprime{$'$}
  \def\cprime{$'$}
\providecommand{\bysame}{\leavevmode\hbox to3em{\hrulefill}\thinspace}
\providecommand{\MR}{\relax\ifhmode\unskip\space\fi MR }
\providecommand{\MRhref}[2]{%
  \href{http://www.ams.org/mathscinet-getitem?mr=#1}{#2}
}
\providecommand{\href}[2]{#2}
\begin{thebibliography}{10}

\bibitem{AMatzner}
S.~Akcay and R.A. Matzner, \emph{{Kerr-de Sitter Universe}}, Class.\ Quantum
  Grav. \textbf{28} (2011), 085012.

\bibitem{CarterKerr}
B.~Carter, \emph{Global structure of the {Kerr} family of gravitational
  fields}, Phys.\ Rev. \textbf{174} (1968), 1559--1571.

\bibitem{Carterseparable}
\bysame, \emph{Hamilton-{J}acobi and {S}chr\"odinger separable solutions of
  {E}instein's equations}, Commun.\ Math.\ Phys. \textbf{10} (1968), 280--310.
  \MR{0239841 (39 \#1198)}

\bibitem{CarterlesHouches}
\bysame, \emph{Black hole equilibrium states}, Black Holes (C.\ de~Witt and B.\
  de~Witt, eds.), Gordon \& Breach, New York, London, Paris, 1973, Proceedings
  of the Les Houches Summer School.

\bibitem{ChANoP}
P.T. Chru\'{s}ciel, \emph{On space-times with {${\rm U}(1)\times {\rm U}(1)$}
  symmetric compact {C}auchy surfaces}, Ann.\ Phys. \textbf{202} (1990),
  100--150. \MR{MR1067565 (91h:83007)}

\bibitem{ChConformalBoundary}
\bysame, \emph{Conformal boundary extensions of {Lorentzian} manifolds}, Jour.\
  Diff.\ Geom. \textbf{84} (2010), 19--44, arXiv:gr-qc/0606101. \MR{2629508
  (2011j:53131)}

\bibitem{CCGP}
P.T. Chru\'{s}ciel, J.~Cortier, and A.~Garcia-Parrado, \emph{{On the global
  structure of the Pomeransky-Senkov black holes}},  \textbf{14} (2009),
  1779--1856, arXiv:0911.0802 [gr-qc]. \MR{2872470 (2012m:83032)}

\bibitem{ChImaxTaubNUT}
P.T. Chru\'{s}ciel and J.~Isenberg, \emph{Non--isometric vacuum extensions of
  vacuum maximal globally hyperbolic space--times}, Phys. Rev. D (3)
  \textbf{48} (1993), 1616--1628. \MR{MR1236815 (94f:83007)}

\bibitem{chrusciel:rendall:SCC}
P.T. Chru{\'s}ciel and A.D. Rendall, \emph{Strong cosmic censorship in vacuum
  space-times with compact, locally homogeneous {C}auchy surfaces}, Ann.
  Physics \textbf{242} (1995), no.~2, 349--385.

\bibitem{CSPS}
P.T. Chru\'{s}ciel and S.J.~Szybka, \emph{Stable causality of the {Pomeransky-Senkov} black holes}, Adv.\ Theor.\ Math.\ Phys. \textbf{15}
  (2011), 175--178, arXiv:1010.0213 [hep-th].

\bibitem{Demianski}
M.~Demia\'nski, \emph{Some new solutions of the {Einstein} equations of
  astrophysical interest}, Acta Astronomica \textbf{23} (1973), 197--231.

\bibitem{EmparanReall}
R.~Emparan and H.S. Reall, \emph{A rotating black ring in five dimensions},
  Phys.\ Rev.\ Lett. \textbf{88} (2002), 101101, arXiv:hep-th/0110260.

\bibitem{EmparanReallLR}
\bysame, \emph{{Black Holes in Higher Dimensions}}, Living Rev.\ Rel.
  \textbf{11} (2008), 6, arXiv:0801.3471 [hep-th].

\bibitem{GibbonsHawkingCEH}
G.W. Gibbons and S.W. Hawking, \emph{Cosmological event horizons,
  thermodynamics, and particle creation}, Phys.\ Rev. \textbf{D15} (1977),
  2738--2751.

\bibitem{GibbonsHerdeiro}
G.W. Gibbons and C.A.R. Herdeiro, \emph{{Supersymmetric rotating black holes
  and causality violation}}, Class.\ Quantum Grav. \textbf{16} (1999),
  3619--3652, arXiv:hep-th/9906098.

\bibitem{gowdy71}
R.H. Gowdy, \emph{Gravitational waves in closed universes}, Phys. Rev. Lett.
  \textbf{27} (1971), 826.

\bibitem{HE}
S.W. Hawking and G.F.R. Ellis, \emph{The large scale structure of space-time},
  Cambridge University Press, Cambridge, 1973, Cambridge Monographs on
  Mathematical Physics, No. 1. \MR{MR0424186 (54 \#12154)}

\bibitem{HT}
M.~Henneaux and C.~Teitelboim, \emph{Asymptotically anti--de {S}itter spaces},
  Commun.\ Math.\ Phys. \textbf{98} (1985), 391--424. \MR{86f:83030}

\bibitem{Misner}
C.W. Misner, \emph{Taub--{N}{U}{T} space as a counterexample to almost
  anything}, Relativity Theory and Astrophysics, AMS, Providence, Rhode Island,
  1967, Lectures in Appl. Math., vol. 8, pp.~160--169.

\bibitem{NUT}
E.~Newman, L.~Tamburino, and T.~Unti, \emph{Empty-space generalization of the
  {S}chwarzschild metric}, Jour.\ Math.\ Phys. \textbf{4} (1963), 915--923.
  \MR{MR0152345 (27 \#2325)}

\bibitem{PS}
A.A. Pomeransky and R.A. Senkov, \emph{{Black ring with two angular momenta}},
  (2006), hep-th/0612005.

\bibitem{RingstromGowdyAtPastInfinity}
H.~Ringstr{\"o}m, \emph{Data at the moment of infinite expansion for polarized
  {G}owdy}, Class.\ Quantum Grav. \textbf{22} (2005), 1647--1653.

\bibitem{RingstroemSCC}
H.~Ringstr{\"o}m, \emph{Strong cosmic censorship in {$T^{3}$-{G}owdy}
  space-times}, Ann. of Math. (2) \textbf{170} (2009), 1181--1240.
  \MR{MR2600872}

\bibitem{Exactsolutions2}
H.~Stephani, D.~Kramer, M.~MacCallum, C.~Hoenselaers, and E.~Herlt, \emph{Exact
  solutions of {E}instein's field equations}, Cambridge Monographs on
  Mathematical Physics, Cambridge University Press, Cambridge, 2003 (2nd ed.).
  \MR{MR2003646 (2004h:83017)}

\bibitem{Taub}
A.H. Taub, \emph{Empty space-times admitting a three parameter group of
  motions}, Ann.\ of Math.\ (2) \textbf{53} (1951), 472--490. \MR{MR0041565
  (12,865b)}

\bibitem{Walker}
M.~Walker, \emph{Bloc diagrams and the extension of timelike two--surfaces},
  Jour.\ Math.\ Phys. \textbf{11} (1970), 2280--2286.

\end{thebibliography}

\end{document}